\documentclass{elsarticle}

\usepackage[margin=2cm]{geometry}
\usepackage{framed,multirow}

\usepackage{amssymb}
\usepackage{amsmath}
\usepackage{latexsym}
\allowdisplaybreaks
\usepackage{tabularx}

\usepackage{graphicx}
\usepackage{rotating, tabstackengine}

\usepackage{url}
\usepackage{xcolor}
\definecolor{newcolor}{rgb}{.8,.349,.1}

\usepackage[colorlinks=true,citecolor=blue]{hyperref}
\usepackage[title]{appendix}
\usepackage{siunitx}
\usepackage{makecell}
\usepackage{comment}
\usepackage[labelfont=bf]{caption}
\usepackage{subcaption}
\usepackage{float}

\DeclareMathAlphabet{\mathcal}{OMS}{cmsy}{m}{n}

\def\snm#1{\textcolor{black}{#1}}

\makeatletter
\def\ps@pprintTitle{%
	\let\@oddhead\@empty
	\let\@evenhead\@empty
	\def\@oddfoot{}%
	\let\@evenfoot\@oddfoot}
\makeatother

\begin{document}


\begin{frontmatter}

\title{A unified multi-phase and multi-material formulation for combustion modelling}%

\author{M. \snm{Nikodemou}\corref{cor1}}
\ead{mn441@cam.ac.uk}
\cortext[cor1]{Corresponding author.}
\author{L. \snm{Michael}}
\ead{lm355@cam.ac.uk}
\author{N. \snm{Nikiforakis}}
\ead{nn10005@cam.ac.uk}

\address{Laboratory for Scientific Computing, Cavendish Laboratory, Department of Physics, University of Cambridge, UK}


\begin{abstract}
	The motivation of this work is to produce an integrated formulation for material response (e.g.\ elastoplastic, viscous, viscoplastic etc.) due to detonation wave loading. Here, we focus on elastoplastic structural response. In particular, we are interested to capture miscible and immiscible behaviour within condensed-phase explosives arising from the co-existence of a reactive carrier mixture of miscible materials, and several material interfaces due to the presence of immiscible impurities such as particles or cavities. The dynamic and thermodynamic evolution of the explosive is communicated to one or more inert confiners through their shared interfaces, which may undergo severe topological change. We also wish to consider elastic and plastic structural response of the confiners, rather than make a hydrodynamic assumption for their behaviour. Previous work by these authors has met these requirements by means of the simultaneous solution of appropriate systems of equations for the behaviour of the condensed-phase explosive and the elastoplastic behaviour of the confiners. To that end, both systems were written in the same mathematical form as a system of inhomogeneous hyperbolic partial differential equations which were solved on the same discrete space using the same algorithms, as opposed to coupling fluid and solid algorithms (co-simulation). In the present work, we employ a single system of partial differential equations (PDEs) proposed by Peshkov and Romenski, which is able to account for different states of matter by means of generalising the concept of distortion tensors beyond solids. We amalgamate that formulation with a single system of PDEs which meets the requirement of co-existing miscible and immiscible explosive mixtures. We present the mathematical derivation and construct appropriate algorithms for its solution. The resulting model is validated against exact solutions for several use-cases, including mechanically- and thermally-induced, inviscid and viscous detonations. Results indicate that the model can accurately simulate a very broad range of problems involving the nonlinear interaction between reactive and inert materials within a single framework. 
\end{abstract}


\end{frontmatter}


\section{Introduction}
\label{sec:intro}
The simultaneous numerical modelling of multiple interacting materials and states of matter is applicable in many areas of physics. This drives research in continuum modelling, with the aim to accurately simulate several interacting processes at the minimum possible computational cost. The motivation behind this work originates from applications in combustion modelling. Examples include the controlled sensitisation of explosives using impurities like air-cavities, glass micro-balloons and solid particles and the modelling of confined detonations, initiated by the impact of a projectile. In both of these scenarios, within the computational domain, the explosive coexists with other materials, which typically have different mechanical properties; for example, an elastoplastic solid confining a liquid explosive. This triggers the need for a model with the capability of simultaneously simulating inert and reactive fluids, as well as viscous fluids or elastoplastic solids. 

There are three main components one needs to consider when developing such an integrated model; the formulation governing reactive flows, the formulation describing structural response, and a method of coupling the two together. Following the authors' previous work, we use the term \textit{structural response} to refer to the elastoplastic deformation of solid materials under loading. The more general term \textit{material response} is also adopted from Peshkov and Romenski \cite{peshkovromenski}, to encapsulate the whole material spectrum from inviscid/viscous fluids to elastoplastic solids. 
 
In this work, we are particularly interested in accurately capturing the reaction zone in combustion applications. We therefore restrict ourselves to reactive formulations that allow for distinct equations of state for the explosive reactants and products. Such formulations involve a numerically mixed zone, in which reactants are converted to products and eventually form the detonation wave. The formulation should also be non-restrictive in the aspect of material response; it should allow for the modelling of both elastic, plastic or elastoplastic solids and inviscid or viscous fluids. It should be noted that we use the term \textit{solid} to refer to materials that undergo structural response, as opposed to solid materials that are hydrodynamically approximated. As far as the coupling method is concerned, we aim to minimise the total number of mathematical systems needed to describe problems of interest. Therefore, our goal is to propose a single formulation suitable for multi-phase (reactive) and multi-material (e.g.\ solid-fluid, solid-solid, fluid-fluid) simulations. 

Formulations describing explosives can usually be classified as either multi-phase models or models of augmented Euler form. The choice between the two is made depending on the physical properties of the explosive material and the application under consideration. 
The most general multi-phase model is the Baer-Nunziato (BN) model \cite{BN}, which has distinct mass, momentum and energy conservation laws for each phase (reactants and products). The volume fraction of one of the phases is advected to distinguish between them and source terms are used to account for exchange processes. Later papers have examined the necessity of total non-equilibrium in various applications and have proposed a number of reduced versions of the BN model, e.g.\ \cite{BN:review, BN:reduced, saurel99}. Formulations of the multi-phase class are mostly suitable for modelling porous solid explosives. 
Formulations of the augmented Euler class (e.g.\ \cite{banks07, banks08}) are, as suggested by the name, augmented forms of the Euler equations; they consist of conservation laws for the mass, momentum and energy and an additional equation for evolving the reactants' mass fraction, for example. Further, equilibrium conditions between materials are also typically required to close the system. This class of reactive models works well for gaseous explosives. 

Michael and Nikiforakis \cite{hybrid} proposed a formulation (henceforth referred to as MiNi16) for the interaction of a {reactant-product} mixture and an inert hydrodynamic material. This is a hybrid, diffuse-interface model which combines the advantages of both classes mentioned above. It uses the augmented Euler model of Banks et al.\ \cite{banks08} to govern the reaction process and the two-phase model of Allaire et al.\ \cite{allaire} for the interface between the explosive and inert materials. The full MiNi16 formulation can be reduced to the aforementioned limiting cases, for example if the explosive is unconfined (reactive model) or if there is no reaction taking place (fluid-fluid model). When comparing the existing reactive models against the requirements of our new formulation, MiNi16 provides both a methodology for simulating reactive flows, but also a desirable framework for interactions between explosive and inert materials, all under a single formulation. 

Solids have traditionally been modelled in the Lagrangian or Arbitrary-Lagrangian-Eulerian (ALE) framework \cite{lagrangian_book, ALE, ALE_hirt}, where the mesh is allowed to deform. A different methodology which bypasses issues of Lagrangian and ALE methods related to large distortions considers the evolution of the deformation tensor, $F$, under the Eulerian frame of reference \cite{godunovromenski, kondaurov, plohr}. A number of high-resolution, shock-capturing methods have been developed for the system's solution, including linearised solvers that capture the full 7-wave configuration in solids \cite{millercolella, barton09, barton10}, an adaptation to the 3-wave HLLC Riemann solver \cite{gavrilyuk}, MUSTA-type fluxes \cite{titarev} and approximate and exact Riemann solvers \cite{barton09, miller}. 
Even though the aforementioned models describe solid mechanics, they differ in the specific assumptions they make. Some formulations describe non-linear elasticity \cite{plohr, barton09, gavrilyuk, titarev, miller}, some include plasticity effects \cite{godunovromenski, kondaurov, millercolella, barton10}, and others include viscoplasticity effects \cite{simo, ortiz, favrie}. There are in general two different approaches of incorporating non-elasticity effects. The first approach involves the multiplicative decomposition of the deformation tensor into elastic and plastic parts, $F = F^e F^p$ (see \cite{kondaurov, millercolella, ortiz}), while the second approach adds a plastic relaxation source term to the evolution equation of $F$ (see \cite{godunovromenski, barton10, favrie}). 

Peshkov and Romenski \cite{peshkovromenski} proposed a hyperbolic formulation (henceforth referred to as Godunov-Peshkov-Romenski, or GPR model) that unifies the two main branches of continuum mechanics; fluid and solid mechanics. This is based on the classification of materials according to the characteristic time, $\tau$, taken for a continuum particle to rearrange with its neighbours, called the particle settled lifetime (PSL). For varying values of $\tau$, the GPR formulation covers the full spectrum of material responses continuously, including elastic and elastoplastic solids as well as viscous and inviscid fluids. The formulation itself is almost identical to the system of Godunov and Romenski \cite{godunovromenski} for elastoplastic solids, but had previously only been applied to solid mechanics. The PSL acts through a possibly stiff relaxation source term in the evolution equation of the distortion tensor. This stiffness, along with the quasi-conservative nature of the model poses a challenge for its numerical solution, which has been the subject of a number of papers, for example \cite{dumbser, boscheri, jackson, jackson19}. When compared against the requirements of our formulation, this unified theory of continuum modelling seems promising. 

Various extensions of the GPR model exist, that are applicable to problems with different physical phenomena. Dumbser et al.\ \cite{dumbser} incorporated hyperbolic heat conduction equations into the system, and applied it to benchmark problems with viscous heat-conducting fluids and elastic solids. Other notable extensions include the incorporation of electro-dynamics \cite{dumbser17}, torsion \cite{GPRtorsion} and damage \cite{gabriel}, the application to non-Newtonian fluids and plastic solids \cite{jackson19}, the general relativistic version of the GPR model \cite{romenskinew} and the two-phase model for solid-fluid mixtures \cite{romenski20}. It is also worth mentioning the extension for combustion modelling by Jackson \cite{jacksonthesis}, who augmented the GPR model with an evolution equation for the reactants' mass fraction, $\lambda$. However, this does not allow for distinct equations of state for the explosion reactants and products, therefore the reaction process is not explicitly resolved. 

Existing methodologies for simulating the interaction between explosive and elastoplastic materials in an Eulerian framework follow an interface tracking approach. The two systems governing reactive flows and structural response are solved on either side of the material interface, resulting in an environment of co-simulation, where different numerical algorithms may be required for each system. Communication between the two materials is achieved by specifying material `boundary' conditions, for example mixed-material Riemann solvers, across the interface. The derivation of these is dependent on the specific systems solved and might prove challenging for more complex constitutive models. The studies by Miller and Colella \cite{millercolella2} and Barton et al.\ \cite{barton11} couple an elastoplastic system with the reactive Euler equations by using a Volume-of-Fluid method for interface reconstruction. The methodologies are applied to explosives approximated by ideal gases and the reaction process is described by a simple rate law or program burn. These are however not adequate for modelling condensed-phase explosives. Schoch et al.\ \cite{schoch} couple the more complex five-equation model of Kapila et al.\ \cite{BN:reduced} for condensed-phase explosives with an elastoplastic model for solids, using the `real' Ghost Fluid Method (GFM) at the interface. Michael and Nikiforakis \cite{michael18} couple the MiNi16 model for condensed-phase explosives with an elastoplastic formulation and Michael et al.\ \cite{michael_fourstates} couple four states of matter (plastma arc, elastoplastic solids, fluids and gases) using a Riemann GFM. Jackson and Nikiforakis \cite{jackson_GFM} also used a Riemann GFM to simulate the interface between two GPR-type materials. Since both sides of the interface are described by the same unified model of continuum mechanics, the approximate Riemann solver is derived once and can be used for interactions between materials with any type of material response. The multi-material formulation uses the extension of the GPR model for reactive flows by Jackson \cite{jacksonthesis}, which, however, does not explicitly resolve the reaction zone. To the best of our knowledge, no diffuse-interface models exist for the interaction between explosive and solid materials. Examples of diffuse-interface solid-fluid and solid-solid models can be found in the literature (e.g.\ \cite{favrie09, favrie12, ndanou, ghaisas17, barton19solidfluid}). 

The motivation of this work is to produce an integrated formulation for elastoplastic structural response due to detonation wave loading, in a diffuse interface framework. A further requirement is that the model is suitable for multi-material (solid-fluid, solid-solid, fluid-fluid) simulations. When compared to existing reactive-solid sharp-interface models (e.g.\ \cite{schoch, michael18}), a diffuse-interface formulation would circumvent the requirement of a complex multi-material Riemann solver for interface communication. Moreover, the ability of a single formulation to simulate a wide range of multi-material applications is promising; the algorithmic optimisation of multiple systems (suitable for interfaces between materials with different properties) is likely to be much more laborious than that of a single system. 

To develop the new formulation, we couple the MiNi16 and GPR models to form a single, unified formulation for multi-phase and multi-material simulations. More specifically, we adopt the framework of MiNi16, hence taking advantage of the embedded diffuse-interface approach for the explosive-inert material interface. The GPR model provides the unified continuum theory that allows materials to exhibit any material response and/or heat conduction (following the extension of Dumbser et al.\ \cite{dumbser}), depending on the application in consideration. The approach taken is to effectively `insert' a GPR-type material in each of the two phases of the MiNi16 framework. 

The resulting formulation can be regarded as a modification of the MiNi16 model. It consists of the same three components: an inert material and a mixture of reactants and products. The main difference of the integrated formulation with MiNi16 is that all components are now allowed to exhibit material response and heat conduction. To this end, the derivation of appropriate mixture rules and equilibrium conditions regarding these added physical processes is necessary. 

Alternatively, the integrated formulation may be regarded as an extension of the GPR model for applications in combustion modelling, that can both capture the reaction zone accurately and allow for the interaction between explosive and inert materials. 

The unified, diffuse-interface nature of the proposed formulation implies that we can choose combinations of material properties that render the model an explosive-solid, explosive-fluid, solid-solid, solid-fluid, fluid-fluid, solids, fluids, or explosives model, accordingly. In a nutshell, we propose a single formulation that, in taking specific limits, can recover several models, some of which have already been validated and extensively used, but largely extends the capabilities of each of these individually. 

The rest of this paper is organised as follows. The two constitutive formulations are given in detail in Section \ref{sect:constitutive}. The proposed formulation is presented in Section \ref{sect:form}, and its limiting models and example applications are summarised in Section \ref{sect:limiting}. The numerical methodology employed for the solution of the model is described in Section \ref{sect:methodology}. Section \ref{sect:validation} presents the validation and evaluation of the new formulation and Section \ref{sect:concl} gives the conclusions of this work. 

\section{Constitutive models}
\label{sect:constitutive}
In this section, we summarise the two models that form the basis of our formulation, to aid with the discussion in the rest of this article. This includes the MiNi16 formulation for simulations of condensed-phase explosives and the GPR model of fluid and solid mechanics. For a complete description of the two models, the reader is referred to \cite{hybrid, dumbser}. 

The conventional notation is used, where $\rho, \mathbf{u}, p, \boldmath{\sigma}, E$ represent density, velocity, pressure, stress and total energy, respectively. 
\subsection{MiNi16 model}
\label{sect:hybrid}
Michael and Nikiforakis \cite{hybrid} proposed a formulation for the simulation of an explosive {reactant-product} mixture, confined by an inert material. This uses the {fluid-mixture} approach of Banks et al.\ \cite{banks08} to model the mixture of explosive and products and the {multi-phase} approach of Allaire et al.\ \cite{allaire} to model the sharp interface between the reactive mixture and the inert confiner. 

Let the inert material be denoted as phase 1, and the reactive mixture as phase 2, composed of the explosive reactants, $\alpha$, and the explosive products, $\beta$. The volume fractions for each phase with respect to the global mixture are denoted by $z^{(1)}, z^{(2)}$ and satisfy the saturation constraint ${z^{(1)} + z^{(2)} = 1}$. The mass fraction, $\lambda$, of reactants with respect to phase 2 is used to describe the reaction progress; $\lambda = 1$ when no reaction has occurred, and $\lambda = 0$ when all reactants have turned into products. Superscripts of the form $^{(l)}$ are used to represent the different components, whereas variables bearing no index correspond to the global mixture. 

The MiNi16 system of equations is given by 
\begin{subequations}
\begin{align}
\frac{\partial }{\partial t}(z^{(1)} \rho^{(1)}) + \nabla \cdotp (z^{(1)} \rho^{(1)} \mathbf{u}) &= 0, \label{eq:hybridmass1cons}\\
\frac{\partial }{\partial t}(z^{(2)} \rho^{(2)}) + \nabla \cdotp (z^{(2)} \rho^{(2)} \mathbf{u}) &= 0, \label{eq:hybridmass2cons} \\
\frac{\partial}{\partial t} (\rho \mathbf{u}) + \nabla \cdotp (\rho \mathbf{u} \otimes \mathbf{u} + p I) &= 0,\label{eq:hybridmomentumcons}\\
\frac{\partial}{\partial t} (\rho E) + \nabla \cdotp (\mathbf{u}(\rho E + p)) &= 0, \label{eq:hybridenergycons}\\
\frac{\partial z^{(1)}}{\partial t} + \mathbf{u} \cdot \nabla z^{(1)} &= 0,  \label{eq:hybridztransport}\\
\frac{\partial}{\partial t} (z^{(2)} \rho^{(2)} \lambda) + \nabla \cdotp (z^{(2)} \rho^{(2)} \lambda \mathbf{u}) &= z^{(2)} \rho^{(2)} \mathcal{K},\label{eq:hybridlambda}
\end{align}
\label{eq:hybrid}
\end{subequations}
where $\mathbf{I}$ is the identity matrix and $\mathcal{K}$ is the rate of conversion from reactants to products. Equations \eqref{eq:hybridmass1cons}-\eqref{eq:hybridmass2cons} are the conservation laws for the individual phases' mass, whereas \eqref{eq:hybridmomentumcons}-\eqref{eq:hybridenergycons} are the total momentum and energy conservation equations, respectively. 
Finally, the transport equation \eqref{eq:hybridztransport} governs the evolution of the volume fraction and \eqref{eq:hybridlambda} is responsible for chemical reaction. The model is not restrictive in terms of the reaction rate law; any general from can be used for $\mathcal{K}$, usually a function of $\rho^{(2)}, \lambda,$ and $p$ or $T$. 

The total energy is given by
\begin{align}
E = \frac{1}{2} \|\mathbf{u}\|^2 + e,
\end{align}
where $e$ is the specific internal energy. 

The system is closed by an equation of state to describe the three components. The MiNi16 model is in general not restrictive in this respect; any equation of state can be used, including those based on tabulated data. However the general {Mie-Gr\"uneisen} form of equation of state is adopted in \cite{hybrid}, 
\begin{align}
\label{eq:hybridmiegrun}
p^{(l)} = p_{ref}^{(l)} + \rho^{(l)} \Gamma^{(l)} (e^{(l)} - e_{ref}^{(l)}), \quad\text{for } l = 1, \alpha, \beta,
\end{align}
where $p_{ref}, e_{ref}$ are reference curves for pressure and energy and $\Gamma$ is the Gr\"uneisen coefficient, all three being functions of density. 

Note that when modelling reactive flows, the reference energy of the products is adjusted to account for the heat transfer between reactants and products,
\begin{align}
e_{ref} = e_{ref} - \mathcal{Q},
\end{align}
where $\mathcal{Q}$ denotes the heat of detonation. 

Velocity and pressure equilibrium is assumed between all three materials, whereas temperature equilibrium is only imposed between reactants and products,
\begin{subequations}
\begin{align}
p^{(1)} = p^{(2)} = p^{(\alpha)} = p^{(\beta)} = p,\\*
\mathbf{u}^{(1)} = \mathbf{u}^{(2)} = \mathbf{u}^{(\alpha)} = \mathbf{u}^{(\beta)} = \mathbf{u},\\*
T^{(\alpha)} = T^{(\beta)} = T^{(2)}. 
\end{align}
\end{subequations}
A set of weighted mixture rules are defined, for the calculation of global variables in the narrow mixing zones:
\begin{subequations}
\begin{alignat}{4}
&\rho &&= z^{(1)} \rho^{(1)} + z^{(2)} \rho^{(2)}, \qquad &&\text{with } \quad \frac{1}{\rho^{(2)}} &&= \frac{\lambda}{\rho^{(\alpha)}} + \frac{1 - \lambda}{\rho^{(\beta)}}, \\*
&\rho e &&= z^{(1)} \rho^{(1)} e^{(1)} + z^{(2)} \rho^{(2)} e^{(2)}, \qquad && \text{with }\quad e^{(2)} &&= \lambda e^{(\alpha)} + (1-\lambda) e^{(\beta)}.
\label{eq:hybridenergymixture}
\end{alignat}
\end{subequations}

The temperature of a material governed by the {Mie-Gr\"uneisen} equation of state is given by 
\begin{equation}
T^{(i)} = T_{ref}^{(i)} \left(\frac{\rho^{(i)}}{\rho_0^{(i)}}\right)^{\Gamma^{(i)}} + \frac{p - p_{ref}^{(i)}}{\rho^{(i)} \Gamma^{(i)} C_{v}^{(i)}}, \quad i = 1, \alpha, \beta,
\end{equation}
where $C_v$ denotes the specific heat at constant volume. The total temperature can be calculated using the mixture rule proposed in \cite{chinnayya}, 
\begin{align}
T = \frac{\sum_{i=1}^2 z^{(i)} \rho^{(i)} C_v^{(i)} T^{(i)}}{\sum_{i =1}^2 z^{(i)} \rho^{(i)} C_v^{(i)}}.
\end{align}
The individual densities of reactants and products, ${\rho^{(\alpha)}, \rho^{(\beta)}}$, are obtained by means of a root finding method applied on the temperature equilibrium condition, as shown by Banks et al.\ \cite{banks07}. 

\subsection{GPR model}
Peshkov and Romenski \cite{peshkovromenski} presented a unified, hyperbolic formulation of continuum mechanics for simulating both viscous flows and elastoplastic solids under finite deformations. This was achieved by introducing the particle settled life time, $\tau$, which describes the time during which a given particle of the material conserves all bonds with its neighbours. This in turn allows for the clear classification of all types of material responses, depending on $\tau$; e.g.\ $\tau = \infty$ for purely elastic deformations, $\tau < \infty$ for viscous fluids and $\tau = 0$ for ideal fluids. Dumbser et al.\ \cite{dumbser} incorporated heat conduction to the GPR model. 

The formulation reads (using Einstein's summation convention)

\begin{subequations}
\begin{align}
\frac{\partial \rho}{\partial t} + \frac{\partial (\rho u_k)}{\partial x_k} &= 0,\label{eq:GPR_rho}\\*
\frac{\partial (\rho u_i)}{\partial t} + \frac{\partial (\rho u_i u_k + p \delta_{ik} - \sigma_{ik})}{\partial x_k} &= 0,\\*
\frac{\partial A_{ik}}{\partial t} + \frac{\partial (A_{im} u_m)}{\partial x_k} + u_j \Big( \frac{\partial A_{ik}}{\partial x_j} - \frac{\partial A_{ij}}{\partial x_k}\Big) &= -\frac{\psi_{ik}}{\theta_1 (\tau_1)},\label{eq:GPR_A}\\*
\frac{\partial (\rho J_i)}{\partial t} + \frac{\partial (\rho J_i u_k + T \delta_{ik})}{\partial x_k} &= -\frac{\rho H_i}{\theta_2 (\tau_2)},\\*
\frac{\partial (\rho s)}{\partial t} + \frac{\partial (\rho s u_k + H_k)}{\partial x_k} &= \frac{\rho \psi_{ik} \psi_{ik}}{\theta_1 (\tau_1) T} + \frac{\rho H_i H_i}{\theta_2 (\tau_2) T}, 
\end{align}\label{eq:GPR}
\end{subequations}

\noindent where $A, \mathbf{J}$ and $s$ denote the material's distortion tensor, thermal impulse vector and entropy, respectively. In this section, we use the notation $E_\rho$ for the partial derivative $\partial E / \partial \rho$, keeping all other state variables constant. 

The variable, $\mathbf{\sigma} = -\rho A^T E_{A}$ is the symmetric viscous shear stress tensor, $T = E_s$ is the temperature, $\tau_1, \tau_2$ denote the strain dissipation and thermal impulse relaxation times, respectively and $\theta_1(\tau_1),\, \theta_2(\tau_2)$ are positive functions. 

The distortion tensor, $A$, characterises the deformation and rotation of a material element. It is a local field, unlike the global deformation tensor commonly used in solids modelling, e.g.\ in \cite{millercolella}. The dissipative source terms are defined as $\mathbf{\psi} = E_A$ and $\mathbf{H} = E_{\mathbf{J}}$, chosen specifically to ensure the entropy inequality. 

It is straightforward to check that the above system also satisfies the conservation of total energy, $E$,
\begin{align}
\frac{\partial (\rho E)}{\partial t} + \frac{\partial (u_k \rho E + u_i (p \delta_{ik} - \sigma_{ik}) + q_k)}{\partial x_k} &= 0,
\end{align}
where $\mathbf{q} = E_s E_{\mathbf{J}}$ is the heat flux. 

The following relation, known as Murnaghan's formula, also holds, 
\begin{align}
\rho &= \rho_0 \det(A). \label{eq:murnaghan}
\end{align}
It can be easily proven that system \eqref{eq:GPR} is overdetermined, as the density PDE \eqref{eq:GPR_rho} can be derived from the distortion tensor evolution equation \eqref{eq:GPR_A}. Therefore, in theory, one could completely discard the continuity equation, and obtain the density from the distortion tensor using Murnaghan's formula \eqref{eq:murnaghan}. However, both equations are included in the system to simplify the numerical implementation of the model and avoid mass conservation errors (see \cite{peshkovromenski} and references therein). 

The equation of state is taken to be a sum of contributions from the microscale, mesoscale and macroscale, 
\begin{align}
E(\rho, s, \mathbf{u}, A, \mathbf{J}) = E_1(\rho, s) + E_2(A, \mathbf{J}) + E_3(\mathbf{u}), 
\end{align}
where $E_1$ and $E_3$ are conventional: they are the usual hydrodynamic internal energy, $e(\rho, s)$, and the specific kinetic energy, $\frac{1}{2} \|\mathbf{u}\|^2$, respectively. 
The mesoscale contribution is chosen to have the form, 
\begin{align}
E_2(A, \mathbf{J}) = \frac{c_s^2}{4} \|\mathrm{dev}(G)\|_F^2+ \frac{c_t^2}{2} \|\mathbf{J}\|^2,
\end{align}
where $\| \mathbf{\cdot} \|_F$ is the Frobenius norm and $\mathrm{dev}(G)$ is the deviator of $G$, both defined by 
\begin{align}
\mathrm{dev}(G) = G - \frac{1}{3} \mathrm{tr}(G) I, \qquad G = A^T A,
\end{align}
and $c_s$ is the characteristic velocity of propagation of transverse perturbations (shear sound speed). The variable $c_t$ is related to the characteristic velocity of heat wave propagation, $c_h$, through
\begin{align}
c_h = \frac{c_t}{\rho} \sqrt{\frac{T}{C_v}}. 
\end{align}

The scalar functions $\theta_1, \theta_2$ are chosen such that the classical Navier-Stokes-Fourier theory is recovered in the limit $\tau_1, \tau_2 \rightarrow 0$ \cite{dumbser}, 
\begin{subequations}
\begin{align}
\theta_1 &= \frac{\tau_1 (c_s)^2}{3 \det(A)^\frac{5}{3}},\\*
\theta_2 &= \tau_2 (c_t)^2 \frac{T_0}{T} \frac{\rho}{\rho_0},
\end{align}
\end{subequations}
where $\rho_0$ and $T_0$ are the reference density and temperature, respectively. 

The relaxation times, $\tau_1, \tau_2$, have the following forms \cite{dumbser, barton11}, 
\begin{subequations}
\begin{align}
\tau_1 &= 
\begin{cases}
\frac{6 \mu}{\rho_0 (c_s)^2}  \qquad &\text{for viscous fluids} \\[3pt]
\tau_0 \Big( \frac{\sigma_0}{\sqrt{\frac{2}{3}}\| \text{dev}(\sigma) \|_F} \Big)^{n} \qquad &\text{for elastoplastic solids}
\end{cases}\\*
\tau_2 &=\frac{\rho_0 \kappa}{T_0 (c_t)^2},
\end{align}
\end{subequations}
where $\mu$ is the dynamic viscosity coefficient, $\tau_0, \sigma_0$ and $n$ are material-specific constants and $\kappa$ is the heat conduction coefficient. 

Having presented the two constitutive models, let us recall the aim of this work; to develop a single formulation suitable for multi-phase and multi-material simulations. As discussed earlier, the diffuse-interface framework embedded within MiNi16, allows us to have both interfaces and reactive mixtures. In addition, the unified theory of continuum modelling of GPR enables us to cover the whole spectrum of material response (including elastoplastic solids and inviscid fluids) in a single model. The aim of this work could therefore be attained by merging the above two models. This forms the subject of the next section. 

\section{Mathematical formulation}
\label{sect:form}
Consider the framework of the MiNi16 model; let phase 1 be the inert material and phase 2 the reactive mixture, composed of reactants $\alpha$ and products $\beta$. The new formulation is derived by assigning a `GPR-type' material to each of the two phases. This introduces the new variables $A^{(1)}, A^{(2)}, \mathbf{J}^{(1)}$ and $\mathbf{J}^{(2)}$ for the phases' distortion tensors and thermal impulse vectors, respectively. 

More specifically, we inherit the treatment of the immiscible interface and the fluid mixture zone from MiNi16 while the GPR model is responsible for the deformation and heat conduction processes. 

The notation convention of the previous section is used. Superscripts of the form $^{(l)}$ are used on physical variables to denote the properties of each component. Variables bearing no index correspond to the global mixture. Einstein's summation convention is also adopted. 

The resulting system of equations is given by 
\begin{subequations}
\begin{align}
\frac{\partial (z^{(1)} \rho^{(1)})}{\partial t} + \frac{\partial (z^{(1)} \rho^{(1)} u_k)}{\partial x_k} &= 0,\label{eq:new_z1rho1}\\*
\frac{\partial (z^{(2)} \rho^{(2)})}{\partial t} + \frac{\partial (z^{(2)} \rho^{(2)} u_k)}{\partial x_k} &= 0,\label{eq:new_z2rho2}\\*
\frac{\partial (\rho u_i)}{\partial t} + \frac{\partial (\rho u_i u_k + p \delta_{ik} - \sigma_{ik})}{\partial x_k} &= 0,\label{eq:new_rhou}\\*
\frac{\partial (z^{(1)})}{\partial t} + \frac{\partial (z^{(1)} u_k)}{\partial x_k} - z^{(1)} \frac{\partial u_k}{\partial x_k} &= 0,\label{eq:new_z1}\\*
\frac{\partial (z^{(2)} \rho^{(2)} \lambda)}{\partial t} + \frac{\partial (z^{(2)} \rho^{(2)} \lambda u_k)}{\partial x_k} &= z^{(2)} \rho^{(2)} \mathcal{K},\label{eq:new_lambda}\\*
\frac{\partial ( \rho E )}{\partial t} + \frac{\partial (\rho E u_k + (p \delta_{ik} - \sigma_{ik}) u_i + q_k)}{\partial x_k} &= 0,\label{eq:new_rhoE}\\*
\frac{\partial A^{(1)}_{ik}}{\partial t} + \frac{\partial (A^{(1)}_{im} u_m)}{\partial x_k} + u_j \Bigg( \frac{\partial A^{(1)}_{ik}}{\partial x_j} - \frac{\partial A^{(1)}_{ij}}{\partial x_k} \Bigg) &= -\frac{\psi^{(1)}_{ik}}{\theta_1^{(1)} \big(\tau_1^{(1)} \big)},\label{eq:new_A1}\\*
\frac{\partial A^{(2)}_{ik}}{\partial t} + \frac{\partial (A^{(2)}_{im} u_m)}{\partial x_k} + u_j \Bigg( \frac{\partial A^{(2)}_{ik}}{\partial x_j} - \frac{\partial A^{(2)}_{ij}}{\partial x_k} \Bigg) &= -\frac{\psi^{(2)}_{ik}}{\theta_1^{(2)} \big(\tau_1^{(2)} \big)},\label{eq:new_A2}\\*
\frac{\partial (\rho J^{(1)}_i)}{\partial t} + \frac{\partial (\rho J^{(1)}_i u_k + T^{(1)} \delta_{ik})}{\partial x_k} &= - \frac{\rho H_i^{(1)}}{\theta^{(1)}_2 \big( \tau_2^{(1)} \big)},\label{eq:new_J1}\\*
\frac{\partial (\rho J^{(2)}_i)}{\partial t} + \frac{\partial (\rho J^{(2)}_i u_k + T^{(2)} \delta_{ik})}{\partial x_k} &= - \frac{\rho H_i^{(2)}}{\theta^{(2)}_2 \big( \tau_2^{(2)} \big)}. \label{eq:new_J2}
\end{align}\label{eq:fullsystem}
\end{subequations}\\[-7pt]
\noindent Equations \eqref{eq:new_z1rho1}-\eqref{eq:new_rhoE} are almost identical to the MiNi16 model, with a few adjustments. The isotropic pressure $p \delta_{ik}$ emerging in the momentum and energy equations of the MiNi16 model is now replaced by the total stress tensor $p \delta_{ik} - \sigma_{ik}$, to allow for shear stresses in viscous or solid materials. An additional heat flux term $q_k$ is also included in the energy equation to account for the process of heat conduction. As a result, the momentum and energy equations \eqref{eq:new_rhou} and \eqref{eq:new_rhoE} are identical to those in the GPR model. Finally, the evolution equations for the new variables $A^{(1)}, A^{(2)}, \mathbf{J}^{(1)}$ and $\mathbf{J}^{(2)}$ are as in the GPR model of Dumbser et al.\ \cite{dumbser}. It is worth noting that the form of equations \eqref{eq:new_J1}-\eqref{eq:new_J2} agree with the hydrodynamic two-phase model of Romenski et al.\ \cite{romenski_heat-conduction}. 

The above system of equations is not sufficient on its own to describe problems of interest. Firstly, the system needs to be closed by an equation of state. Furthermore, following a diffuse-interface framework, the formulation needs to be augmented with mixtures rules and equilibrium conditions relating the global variables to those of individual components. As far as the conventional hydrodynamic variables are concerned, mixture rules can be inherited directly from the MiNi16 model. However, the variables describing material deformation, $A$, and heat conduction, $\mathbf{J}$, are attributes native to the GPR model, but not MiNi16. This creates the need to define physically appropriate mixture rules for these new variables. Finally, the constitutive functions and source terms are to be defined in an appropriate manner. 

\subsection{Constitutive functions and source terms}
Even if we have effectively assigned a `GPR-type' material to phases 1 and 2, the framework of the model still consists of three independent components. Therefore, all variable definitions are inherited from the GPR model and are valid for components $1, \alpha$ and $\beta$. The link between individual variables for $\alpha, \beta$ and mixture variables for phase 2 is achieved by means of the equilibrium conditions and mixture rules defined in subsequent sections. 
We therefore have the following definitions for $l = 1, \alpha, \beta$, 

\begin{subequations}
\begin{align}
p^{(l)} &= (\rho^{(l)})^2 \frac{\partial E^{(l)}}{\partial \rho^{(l)}},\\
\sigma^{(l)} &= - \rho^{(l)} (A^{(l)})^T \frac{\partial E^{(l)}}{\partial A^{(l)}},\\
T^{(l)} &= \frac{\partial E^{(l)}}{\partial s^{(l)}},\\
\mathbf{q}^{(l)} &= T^{(l)} \frac{\partial E^{(l)}}{\partial \mathbf{J}^{(l)}},\\
\psi^{(l)} &= \frac{\partial E^{(l)}}{\partial A^{(l)}},\\
\mathbf{H}^{(l)} &= \frac{\partial E^{(l)}}{\partial \mathbf{J}^{(l)}}.
\end{align}
\end{subequations}
The scalar functions $\theta_1^{(l)}, \theta_2^{(l)}$ for each component $ l = 1, \alpha, \beta,$ are chosen such that the classical Navier-Stokes-Fourier theory is recovered in the limit $\tau_1, \tau_2 \rightarrow 0$ \cite{dumbser},
\begin{subequations} 
\begin{align}
\theta_1^{(l)} &= \frac{\tau_1^{(l)} (c_s^{(l)})^2}{3 \det(A^{(l)})^\frac{5}{3}},\\*
\theta_2^{(l)} &= \tau_2^{(l)} (c_t^{(l)})^2 \frac{ T_0^{(l)}}{T^{(l)}} \frac{\rho}{\rho_0^{(l)}},
\end{align}
\end{subequations}
where $\rho_0^{(l)}, T_0^{(l)}$ are the reference density and temperature, respectively. The relaxation times, $\tau_1^{(l)}, \tau_2^{(l)}$, take the following forms \cite{dumbser, barton11}, 
\begin{subequations}
\begin{align}
\tau_1^{(l)} &= 
\begin{cases}
\frac{6 \mu^{(l)}}{\rho_0^{(l)} (c_s^{(l)})^2}  \qquad &\text{for viscous fluids} \\[3pt]
\tau_0^{(l)} \Big( \frac{\sigma_0^{(l)}}{\sqrt{\frac{2}{3}}\| \text{dev}(\sigma^{(l)} \|_F} \Big)^{n^{(l)}} \qquad &\text{for elastoplastic solids}
\end{cases}\label{eq:tau1}\\*
\tau_2^{(l)} &=\frac{\rho_0^{(l)} \kappa^{(l)}}{T_0^{(l)} (c_t^{(l)})^2},
\end{align}
\end{subequations}
where $\mu^{(l)}$ is the dynamic viscosity coefficient, $\tau_0^{(l)}, \sigma_0^{(l)}$ and $n^{(l)}$ are material specific constants and $\kappa^{(l)}$ is the heat conduction coefficient. 

\subsection{Equilibrium Conditions}
The model presented above makes a number of assumptions. Firstly, velocity equilibrium is assumed between all three components, following the assumptions of the MiNi16 model, 
\begin{align}
\mathbf{u}^{(1)} = \mathbf{u}^{(2)} = \mathbf{u}^{(\alpha)} = \mathbf{u}^{(\beta)} = \mathbf{u}.
\end{align}
As in the MiNi16 model, pressure equilibrium is taken between all components,
\begin{align}	
p^{(1)} = p^{(2)} = p^{(\alpha)} = p^{(\beta)} = p.\label{eq:pequil}
\end{align}
Contrary to the MiNi16 model, where temperature equilibrium is assumed between reactants and products, our model assumes a constant density ratio
\begin{align}
\frac{\rho^{(\alpha)}}{\rho^{(\beta)}} = \frac{\rho^{(\alpha)}_0}{\rho^{(\beta)}_0} \equiv r,\label{eq:constdensratio}
\end{align}
where $\rho^{(\alpha)}_0$ and $ \rho^{(\beta)}_0$ are reference densities for the two materials. This choice is based on the fact that the simple constant density ratio condition avoids the need for a root-finding procedure to obtain the individual densities. As remarked by Stewart et al \cite{stewart}, this closure condition is more suitable when the reaction rate law is mechanical, rather than temperature sensitive, therefore it should be used with caution. In cases where a temperature dependent reaction rate law is required, it is preferable to use the temperature equilibrium condition of MiNi16. 

The distortion tensor is an attribute present in the single-material GPR model, but not in the multi-phase MiNi16 model. The assumptions we make follow the diffuse-interface solid model of Barton \cite{barton19solidfluid}, 
\begin{align}
\text{dev}(-\frac{1}{2}\ln(G^{(\alpha)})) = \text{dev}(-\frac{1}{2}\ln(G^{(\beta)}))
\end{align}
where $G^{(l)} = (A^{(l)})^T A^{(l)}$ is the local quantity corresponding to the Finger tensor. This in turn implies
\begin{align}
\left( \frac{\rho^{(\alpha)}}{\rho_0^{(\alpha)}}\right)^{\frac{2}{3}}(G^{(\alpha)})^{-1} = \left( \frac{\rho^{(\beta)}}{\rho_0^{(\beta)}}\right)^{\frac{2}{3}}(G^{(\beta)})^{-1},
\end{align}	
which further reduces to 
\begin{align}
	G^{(\alpha)} = G^{(\beta)} = G^{(2)}, 
\end{align}
after substitution of the constant density ratio assumption \eqref{eq:constdensratio}. 

Finally, a common thermal impulse vector is assumed between the reactants and products, 
\begin{align}
\mathbf{J}^{(\alpha)} = \mathbf{J}^{(\beta)} = \mathbf{J}^{(2)}. \label{eq:J_equil}
\end{align} 

\subsection{Equation of state}
To close the system, one needs to define the equation of state describing each material. We follow the approach of the GPR model, assuming the decomposition of the total energy into a kinetic, a hydrodynamic, a viscous and a thermal part, for the three components $l = 1, \alpha, \beta$,
\begin{align}
E^{(l)} = E_k^{(l)}(\mathbf{u}) + E_h^{(l)}(\rho^{(l)}, p^{(l)}_{h}) + E_v^{(l)}(A^{(l)}, \rho^{(l)}) + E_t^{(l)}(\mathbf{J}^{(l)}, \rho^{(l)}),
\end{align}
respectively. The internal energy is defined for each component by
\begin{align}
e^{(l)} = E^{(l)} - E_k^{(l)} = E_h^{(l)} + E_v^{(l)} + E_t^{(l)}.
\end{align}
The specific kinetic energy is, as usual, given by
\begin{align}
E_k^{(l)}(\mathbf{u}) = \frac{1}{2} \|\mathbf{u}\|^2.
\end{align}
The hydrodynamic part, $E_h^{(l)}$ can take any classical equation of state. In this work, we use the general Mie-Gr{\"u}neisen family of equations of state, 
\begin{align}
E_h^{(l)} = e_{ref}^{(l)}(\rho^{(l)}) + \frac{p_h^{(l)} - p_{ref}^{(l)}(\rho^{(l)})}{\rho^{(l)} \Gamma^{(l)}(\rho^{(l)})},
\label{eq:MieGruneisen}
\end{align}
with temperature 
\begin{align}
	T^{(l)} = T_{ref}^{(l)}(\rho^{(l)}) \left(\frac{\rho^{(l)}}{\rho_0^{(l)}}\right)^{\Gamma^{(l)}(\rho^{(l)})} + \frac{E_h^{(l)} - e_{ref}^{(l)}(\rho^{(l)})}{C_v^{(l)}}.
\end{align}
In reactive flows, the detonation products carry an additional term in their equation of state function $e_{ref}$, 
\begin{align}
e_{ref} = e_{ref} - \mathcal{Q},
\end{align}
where $\mathcal{Q}$ is the heat of combustion. This accounts for the additional energy released by the combustion of the explosive. 

Note that the hydrodynamic energy is a function of the hydrodynamic pressure, $p_h^{(l)}$, which does not in general coincide with the total pressure, $p^{(l)}$ (see Peshkov and Romenski \cite{peshkovromenski}). 

Finally, the viscous and thermal parts of the energy are defined by 
\begin{align}
E_v^{(l)} &= \frac{c_s^{(l)}(\rho^{(l)})^2}{4} \|{\text{dev}(G^{(l)})}\|^2_F,\label{eq:EOS_vicsous}\\*
E_t^{(l)} &= \frac{c_t^{(l)}(\rho^{(l)})^2}{2} \|\mathbf{J}\|^2,
\end{align}
where $c_s^{(l)}$ and $c_t^{(l)}$ are defined as before, with 
\begin{align}
c_t^{(l)} = c_h^{(l)} \rho^{(l)} \sqrt{\frac{C_v^{(l)}}{T^{(l)}}}. 
\end{align}
In this work, we make the assumption that $c_t$ is a constant, following Jackson \cite{jackson}. 

Given this form for the equations of state, we can evaluate the following constitutive functions, for each component $l = 1, \alpha, \beta$,
\begin{subequations}
\begin{align}
\psi^{(l)} &= (c_s^{(l)})^2 A^{(l)} \text{dev}(G^{(l)}),\label{eq:psi}\\[3pt]
\mathbf{H}^{(l)} &= (c_t^{(l)})^2 \mathbf{J}^{(l)},\label{eq:H}\\[3pt]
\sigma^{(l)} &= - \rho^{(l)} (c_s^{(l)})^2 G^{(l)} \text{dev}(G^{(l)}),\\[3pt]
\mathbf{q}^{(l)} &= (c_t^{(l)})^2 T^{(l)} \mathbf{J}^{(l)}.
\end{align}
\end{subequations}
Further, the pressure for each component can be split into a hydrodynamic and a viscous part, $p = p_{h} + p_{v},$ (superscripts are dropped for simplicity) with
\begin{subequations}
\begin{align}
p_h &= \rho^2 \frac{\partial E_h}{\partial \rho} = p_{ref} + \rho \Gamma (E_h - E_{ref}),\\*
p_v &= \rho^2 \frac{\partial E_v}{\partial \rho} = \frac{1}{2} \rho^2 c_s \frac{dc_s}{d\rho} \|\text{dev}{G}\|^2_F. 
\end{align}
\end{subequations}

\subsection{Mixture rules}
A set of physically sensible mixture rules need to be defined for the artificial mixing zones of the different components. The mixture density and total energy are given by
\begin{align}
\rho &= z^{(1)} \rho^{(1)} + z^{(2)} \rho^{(2)}, &&\text{with} \qquad \frac{1}{\rho^{(2)}} = \frac{\lambda}{\rho^{(\alpha)}} + \frac{(1 - \lambda)}{\rho^{(\beta)}},\label{eq:rho2mixture}\\*
\rho E &= z^{(1)} \rho^{(1)} E^{(1)} + z^{(2)} \rho^{(2)} E^{(2)}, &&\text{with} \qquad E^{(2)} = \lambda E^{(\alpha)} + (1 - \lambda) E^{(\beta)}\label{eq:E2mixture},
\end{align}
as in the MiNi16 model. 
The viscous part of the stress tensor is volume weighted \cite{favrie09}, 
\begin{align}
\sigma &= z^{(1)} \sigma^{(1)} + z^{(2)} \sigma^{(2)}, \qquad \text{with} \qquad \sigma^{(2)} = - \rho^{(2)} (A^{(2)})^\intercal \frac{\partial E^{(2)}}{\partial A^{(2)}}.
\end{align}
Using equation \eqref{eq:E2mixture}, we can evaluate $\sigma^{(2)}$,
\begin{align}
\sigma^{(2)} &= - \rho^{(2)} \left( \lambda (c_s^{(\alpha)})^2 + (1 - \lambda) (c_s^{(\beta)})^2 \right) G^{(2)} \text{dev}(G^{(2)}) = \rho^{(2)}\Bigg( \frac{\lambda}{\rho^{(\alpha)}} \sigma^{(\alpha)} + \frac{(1-\lambda)}{\rho^{(\beta)}} \sigma^{(\beta)} \Bigg),
\end{align}
which is itself in the form $\sigma^{(2)} = - \rho^{(2)} (c_s^{(2)})^2 G^{(2)} \text{dev} G^{(2)} $, if we take the following mixture rule for $ (c_s^{(2)})^2$:
\begin{align}
(c_s^{(2)})^2 &= \lambda (c_s^{(\alpha)})^2 + (1 - \lambda) (c_s^{(\beta)})^2.\label{eq:c_s^2}
\end{align}
The mixture heat flux is mass weighted, following the approach of Romenski et al.\ \cite{romenski07}, 
\begin{align}
\mathbf{q} &= \frac{z^{(1)} \rho^{(1)}}{\rho} \mathbf{q}^{(1)} + \frac{z^{(2)} \rho^{(2)}}{\rho} \mathbf{q}^{(2)}, \qquad \text{with} \qquad \mathbf{q}^{(2)} = T^{(2)} \frac{\partial E^{(2)}}{\partial \mathbf{J}^{(2)}}.
\end{align}
We use the energy mixture rule \eqref{eq:E2mixture} to calculate the heat flux for phase 2, 
\begin{align}
	\mathbf{q}^{(2)} = \left(\lambda(c_t^{(\alpha)})^2 + (1-\lambda) (c_t^{(\beta)})^2\right) T^{(2)} \mathbf{J}^{(2)},
\end{align}
which is itself in the form $(c_t^{(2)})^2 T^{(2)} \mathbf{J}^{(2)}$, if we let the mixture $(c_t^{(2)})^2$ be defined by
\begin{align}
(c_t^{(2)})^2 &= \left(\lambda (c_t^{(\alpha)})^2 + (1-\lambda) (c_t^{(\beta)})^2 \right). \label{eq:c_t^2}
\end{align}
In order to calculate the temperature for phase 2, $T^{(2)} = \frac{\partial E^{(2)}}{\partial s^{(2)}}$, we use the pressure equilibrium condition \eqref{eq:pequil}, along with a mass weighted entropy, 
\begin{align}
s^{(2)} &= \lambda s^{(\alpha)} + (1-\lambda) s^{(\beta)},
\end{align}
which yields 
\begin{align}
	T^{(2)} &= \frac{\lambda\rho^{(\beta)} \Gamma^{(\beta)}T^{(\beta)}T^{(\alpha)} + (1-\lambda) \rho^{(\alpha)} \Gamma^{(\alpha)} T^{(\alpha)} T^{(\beta)}}{\lambda \rho^{(\beta)} \Gamma^{(\beta)} T^{(\beta)} + (1-\lambda) \rho^{(\alpha)} \Gamma^{(\alpha)} T^{(\alpha)}}. \label{eq:T2mixture}
\end{align}
The source functions $\psi^{(2)}$ and $\mathbf{H}^{(2)}$ are calculated using formulae \eqref{eq:psi} and \eqref{eq:H}, along with the derived mixture rules \eqref{eq:c_s^2} and \eqref{eq:c_t^2} for $c_s^{(2)}$ and $c_t^{(2)}$, respectively. It is clear that we also need mixing rules for the strain dissipation time, $\tau_1^{(2)}$, and the thermal impulse relaxation time, $\tau_2^{(2)}$, of the fluid mixture, so that the system's source terms can be evaluated. Since there is no physically correct and established way to do this, we follow the approach of Ghaisas et al.\ \cite{ghaisas17}, to weigh them by volume fractions, 
\begin{align}
\tau_n^{(2)} = \rho^{(2)} \Bigg( \frac{\lambda}{\rho^{(\alpha)}} \tau_n^{(\alpha)} + \frac{(1-\lambda)}{\rho^{(\beta)}} \tau_n^{(\beta)} \Bigg), \qquad n = 1, 2.
\end{align}

One of the most challenging aspects of developing diffuse-interface models is the physically appropriate definition of mixture rules for different variables in the artificial mixing zones. The mixture rules summarised in this section have been tested and found to perform well under different applications. 

\section{Limiting models}
\label{sect:limiting}
\begin{figure}[!ht]
	\centering
	\includegraphics[width=\linewidth]{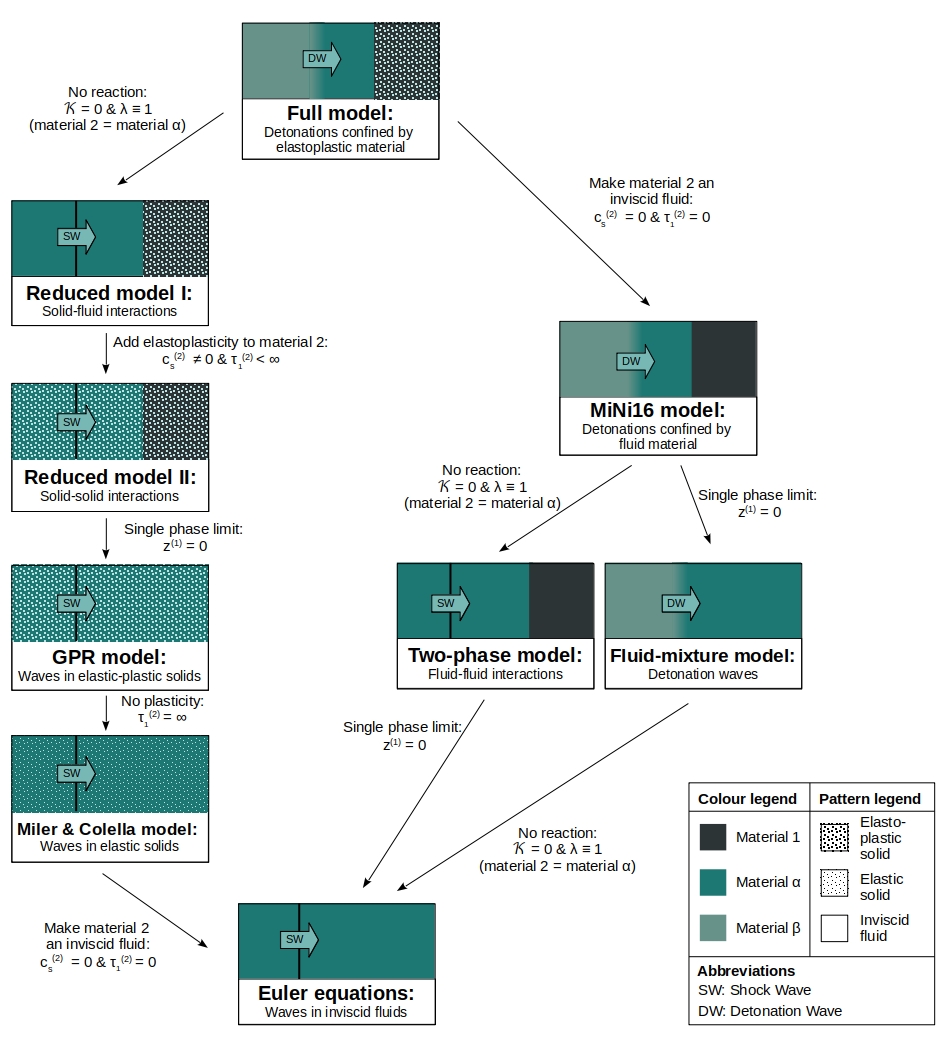}
	\caption{Illustration of how the full model is simplified to its limiting cases and reduced models. Examples of possible applications are also given in each case by means of a schematic diagram. This diagram is not exhaustive; the full spectrum of material response and heat conduction have not been illustrated. }
	\label{fig:limiting}
\end{figure}
The new formulation is built upon the merging of two existing formulations, namely the MiNi16 model \cite{hybrid} for condensed-phase explosives and the unified GPR model \cite{dumbser}. Both constitutive formulations can therefore be regarded as limiting models of our proposed model. More specifically, we recover:
\begin{itemize}
	\item the MiNi16 model in the limit of no deformation or heat conduction, i.e.\ if $c_s^{(l)} = 0$ and $c_t^{(l)} = 0$ for $l = 1,2$
	\item the GPR model in the limit of a single inert material, i.e.\ as $z^{(2)} \rightarrow 0$
\end{itemize}
Each of these limiting models also reduces to simpler models when appropriate limits are taken. For example, MiNi16 reduces to
\begin{itemize}
	\item the fluid-mixture model of Banks et al.\ \cite{banks08} for combustion modelling in the limit of a pure explosive material, i.e.\ $z^{(1)} \rightarrow 0$,
	\item the two-phase model of Allaire et al.\ \cite{allaire} for inert, immiscible, flows, in the limit of no chemical reaction, i.e.\ if $\lambda = 0$ or $1$
	\item the single-component Euler equations in the limit of a single inert material, i.e.\ as $z^{(2)} \rightarrow 0$
\end{itemize}
The GPR model reduces to 
\begin{itemize}
	\item the model of non-linear elasticity by Miller and Collela \cite{millercolella} (with no plasticity source terms) in the limit of no heat conduction, i.e.\ if $c_t = 0$ and $\tau_1 \rightarrow \infty$
\end{itemize}

Figure \ref{fig:limiting} illustrates how the full model is simplified to its limiting cases and reduced models. Simple example applications are given for each model by means of a schematic diagram. The application presented for the full system is based on a one-dimensional analogue to a confined rate stick problem, and is reduced to simpler Riemann problems for each effective model. It should be noted that the diagram in Figure \ref{fig:limiting} is by no means exhaustive. If we were to also consider the process of heat conduction and the full spectrum of material response (e.g.\ Newtonian and non-Newtonian viscosity), the same diagram would become quite crowded. Instead, the diagram focuses on one-dimensional applications involving combinations of fluid explosives and elastoplastic solids to complement the majority of the validation tests mentioned here. 

\section{Numerical Methodology}
\label{sect:methodology}
The full system of equations \eqref{eq:fullsystem} can be written in the form 
\begin{align}
\frac{\partial \mathbf{U}}{\partial t} + \nabla \cdot \mathbf{F}(\mathbf{U}) + \mathbf{B}(\mathbf{U}) \cdot \nabla (\mathbf{U}) = \mathbf{S}(\mathbf{U}).\label{eq:system}
\end{align}
For its solution, we adopt the finite volume framework. We split the system into two sub-systems, to be solved independently, 
\begin{align}
\frac{\partial \mathbf{U}}{\partial t} + \nabla \cdot \mathbf{F}\left(\mathbf{U}\right) + \mathbf{B}\left(\mathbf{U}\right) \cdot \nabla \mathbf{U} &= \mathbf{0},\label{eq:homogeneous}\\*
\frac{\partial \mathbf{U}}{\partial t} &= \mathbf{S}\left(\mathbf{U}\right).\label{eq:source} 
\end{align}
Denote by $\mathcal{H}^{\Delta t}, \mathcal{S}^{\Delta t}$ the operators solving the homogeneous hyperbolic system \eqref{eq:homogeneous} and the ordinary differential equation (ODE) \eqref{eq:source}, respectively. Then, Strang's splitting is adopted, to advance the solution of \eqref{eq:system} for one time step,
\begin{align}
\mathbf{U}^{n+1} = \mathcal{S}^{\frac{\Delta t}{2}} \mathcal{H}^{\Delta t} \mathcal{S}^{\frac{\Delta t}{2}} \mathbf{U}^n.\label{eq:strang}
\end{align}

\subsection{Homogeneous system}
For the homogeneous system \eqref{eq:homogeneous}, we split the variables in two categories; the conservative (where the corresponding row of $\mathbf{B}$ is zero) and the non-conservative. It is easily seen from the system of equations \eqref{eq:fullsystem} that all variables are conservative except from the distortion tensors $A^{(l)}$ and the volume fraction $z^{(1)}$. With no loss of generality and for the sake of simplicity, let us consider spatial evolution in the $x-$direction. The conserved variables are evolved using the conventional conservative update formula
\begin{align}
\mathbf{U}^{n+1}_i = \mathbf{U}^{n}_i + \frac{\Delta t}{\Delta x} \left( \mathbf{F}_{i-\frac{1}{2}} - \mathbf{F}_{i + \frac{1}{2}}\right).
\end{align}
The non-conservative variables are integrated using the mid-point rule, following the method of Johnsen and Colonius \cite{johnsen},
\begin{subequations}
\begin{align} 
\int_{x_{i-\frac{1}{2}}}^{x_{i+\frac{1}{2}}} z^{(1)} \frac{\partial u_1}{\partial x} dx  &\approx \frac{z^{(1)}}{\Delta x} \left( (u_1)_{i+\frac{1}{2}} - (u_1)_{i-\frac{1}{2}}\right),\label{eq:non_cons1}\\*
\int_{x_{i-\frac{1}{2}}}^{x_{i+\frac{1}{2}}} u_1 \frac{\partial A^{(l)}_{jk}}{\partial x} dx & \approx \frac{(u_1)_i}{\Delta x} \left((A^{(l)}_{jk})_{i+\frac{1}{2}} - (A^{(l)}_{jk})_{i-\frac{1}{2}} \right),\\*
\int_{x_{i-\frac{1}{2}}}^{x_{i+\frac{1}{2}}} u_k \frac{\partial A^{(l)}_{jk}}{\partial x} dx & \approx \frac{(u_k)_i}{\Delta x} \left((A^{(l)}_{jk})_{i+\frac{1}{2}} - (A^{(l)}_{jk})_{i-\frac{1}{2}} \right),\label{eq:non_cons3}
\end{align}
\end{subequations}
where the boundary values are taken to be those from the appropriate Riemann solver. 
Expressions \eqref{eq:non_cons1}-\eqref{eq:non_cons3} are added on to the conventional conservative formula to give the corrected update for non-conservative variables. It now remains to develop an approximate Riemann solver for our system. This will provide us both with the intercell fluxes, $\mathbf{F}_{i+\frac{1}{2}}$, and the intermediate states $\mathbf{U}_{i+\frac{1}{2}}$ needed for the non-conservative updates. 

\subsubsection{HLLD solver}
\begin{figure}
	\centering
	\includegraphics[width=0.4\linewidth]{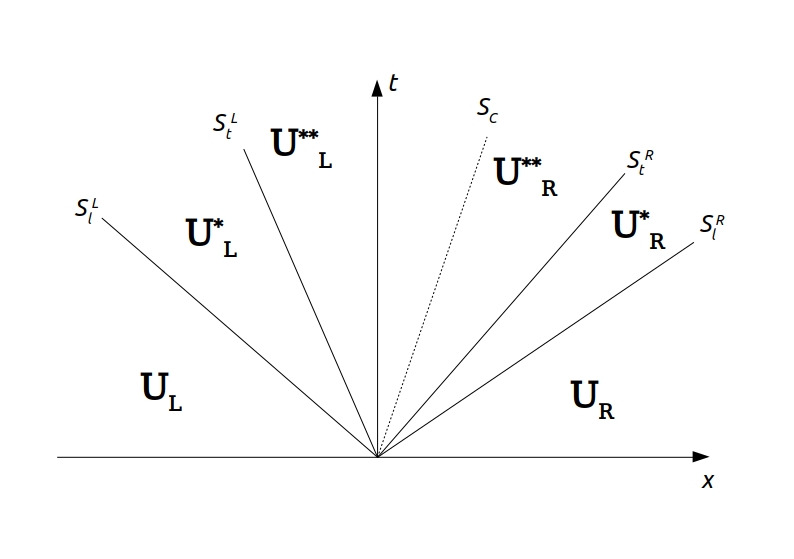}
	\caption{Illustration of the Riemann fan approximated by the HLLD solver}
	\label{fig:HLLD_fan}
\end{figure}
The approximate solution to the Riemann problem is found by adapting the HLLD solver from \cite{ortega}. Depending on the initial conditions, the Riemann problem solution can consist of up to seven distinct waves separating up to eight constant states. Moving from outside towards the centre, the waves are: two outer longitudinal waves followed by two pairs of transverse shear waves and finally the contact wave. However, to simplify the problem, approximating the two pairs of transverse waves as single transverse shocks has been found to be sufficient \cite{barton19solidfluid}. The resulting wave configuration and name convention for the wave speeds and intermediate states are shown in Figure \ref{fig:HLLD_fan}. 

The solution for the inner star states is obtained by applying the Rankine-Hugoniot conditions across the four outermost waves,
\begin{subequations}
\begin{align} 
S_l^K (\mathbf{U}^{*K} - \mathbf{U}^K) &= (\mathbf{F}(\mathbf{U}^{*K}) - \mathbf{F}(\mathbf{U}^K))\\*
S_t^K (\mathbf{U}^{**K} - \mathbf{U}^{*K}) &= (\mathbf{F}(\mathbf{U}^{**K}) - \mathbf{F}(\mathbf{U}^{*K})),
\end{align}
\end{subequations}
where the superscript $K$ will be used to denote left $L$ or right $R$ variables. 

Following the work of Barton \cite{barton19solidfluid} and treating the thermal impulse vector as a thermal analogue of momentum, we assume the following:
\begin{itemize}
	\item Normal components of velocity, total stress, thermal impulse and temperature can jump across longitudinal waves but are constant across shear waves and the contact. 
	\item Transverse components of velocity, total stress and thermal impulse can jump across transverse waves and the contact, but remain constant across longitudinal waves. 
\end{itemize}
We present the Riemann solution across the first dimension. The constant states behind the longitudinal waves are given by
\begin{align}
\hspace{-20pt}\begin{pmatrix} 
z^{(1)} \rho^{(1)} 
\\ z^{(2)} \rho^{(2)} 
\\ \Bigl( \begin{smallmatrix} 
\rho u_1 \\ 
\rho u_2 \\ 
\rho u_3 
\end{smallmatrix}\Bigr) \\ 
z^{(1)} \\ 
z^{(2)} \rho^{(2)} \lambda \\ 
\rho E  \\ 
\Bigl( \begin{smallmatrix} 
A_{11} & A_{12} & A_{13} \\ 
A_{21} & A_{22} & A_{23} \\ 
A_{31} & A_{32} & A_{33} 
\end{smallmatrix} \Bigr)^{(1)} \\ 
\Bigl( \begin{smallmatrix} 
A_{11} & A_{12} & A_{13} 
\\ A_{21} & A_{22} & A_{23} 
\\ A_{31} & A_{32} & A_{33} 
\end{smallmatrix} \Bigr)^{(2)} \\ 
\Bigl( \begin{smallmatrix} 
\rho J_1 \\ 
\rho J_2 \\ 
\rho J_3 
\end{smallmatrix} \Bigr)^{(1)} \\ 
\Bigl( \begin{smallmatrix} 
\rho J_1 \\ 
\rho J_2 \\ 
\rho J_3 
\end{smallmatrix} \Bigr)^{(2)}
\end{pmatrix}^{*K} \hspace{-12pt} &= \hspace{-2pt}\frac{(u_1)^K - S_l^K}{S_c - S_l^K} \begin{pmatrix} 
z^{(1)} \rho^{(1)} \\ 
z^{(2)} \rho^{(2)} \\ 
\Bigl( \begin{smallmatrix} \rho S_c \\ 
\rho u_2 \\ 
\rho u_3 \end{smallmatrix}\Bigr) \\ 
z^{(1)} (S_c - S_l) / (u_1 - S_l) \\ 
z^{(2)} \rho^{(2)} \lambda \\ 
\rho E  \\ 
\Bigl( \begin{smallmatrix} 
A_{11} & 0 & 0 \\ 
A_{21} & 0 & 0 \\ 
A_{31} & 0 & 0 
\end{smallmatrix} \Bigr)^{(1)} \\ 
\Bigl( \begin{smallmatrix} 
A_{11} & 0 & 0 \\ 
A_{21} & 0 & 0 \\ 
A_{31} & 0 & 0 
\end{smallmatrix} \Bigr)^{(2)} \\ 
\Bigl( \begin{smallmatrix} 
\rho J_c \\ 
\rho J_2 \\ 
\rho J_3 \end{smallmatrix} \Bigr)^{(1)} \\ 
\Bigl( \begin{smallmatrix} 
\rho J_c \\ 
\rho J_2 \\ 
\rho J_3 \end{smallmatrix} \Bigr)^{(2)}
\end{pmatrix}^K \hspace{-12pt} + \frac{1}{S_c - S_l^K} 
\begin{pmatrix} 
0 \\ 
0 \\ 
\Bigl(\begin{smallmatrix} 
0\\
0\\
0\end{smallmatrix} \Bigr) \\ 
0 \\ 
0 \\ 
q_1 - q_1^* + (S_c - u_1) \left[\rho S_c (u_1 - S_l) - (p - \sigma_{11})\right] \\ 
(S_c - S_l) \Bigl( \begin{smallmatrix} 
0 & A_{12} & A_{13} \\ 
0 & A_{22} & A_{23} \\ 
0 & A_{32} & A_{33} \end{smallmatrix}\Bigr)^{(1)} \\ 
(S_c - S_l) \Bigl( \begin{smallmatrix} 
0 & A_{12} & A_{13} \\ 
0 & A_{22} & A_{23} \\ 
0 & A_{32} & A_{33} \end{smallmatrix}\Bigr)^{(2)} \\ 
\Bigl( \begin{smallmatrix} 
0 \\ 
0 \\ 
0 \end{smallmatrix} \Bigr)^{(1)} \\
\Bigl( \begin{smallmatrix} 
0 \\ 
0 \\ 
0 \end{smallmatrix} \Bigr)^{(2)} \end{pmatrix}^K
\end{align}
The normal component of the total stress tensor behind the longitudinal waves is given by 
\begin{align}
p^{*K} - \sigma_{11}^{*K} &= p^{K} - \sigma_{11}^{K} + \rho^K \left(u_1^K - S_c\right)\left(u_1^K - S_l^K\right),
\end{align}
whereas the temperatures and heat flux are
\begin{align}
T^{(n)*K} &= T^{(n)K} + \rho^K\left(J_1^{(n)K} - J^{(n)}_c\right)\left(u_1^K - S_l^K\right), \qquad \text{for} \quad n = 1, 2\\*
q_1^{*K} &= \left(\frac{z^{(1)} \rho^{(1)}}{\rho}\right)^K (c_t^{(1)})^2 T^{(1)*K} J_c^{(1)} + \left(\frac{z^{(2)} \rho^{(2)}}{\rho}\right)^K \left(\lambda^K (c_t^{(\alpha)})^2  + (1 - \lambda^K) (c_t^{(\beta)})^2\right) T^{(2)*K} J_c^{(2)}.
\end{align}
The intermediate states behind the transverse waves are given by
\begin{align}
\begin{pmatrix} 
z^{(1)} \rho^{(1)} \\ z^{(2)} \rho^{(2)} \\ \Bigl( \begin{smallmatrix} \rho u_1 \\ \rho u_2 \\ \rho u_3 \end{smallmatrix}\Bigr) \\ z^{(1)} \\ z^{(2)} \rho^{(2)} \lambda \\ \rho E  \\ \Bigl( \begin{smallmatrix} A_{11} & A_{12} & A_{13} \\ A_{21} & A_{22} & A_{23} \\ A_{31} & A_{32} & A_{33} \end{smallmatrix} \Bigr)^{(1)} \\ \Bigl( \begin{smallmatrix} A_{11} & A_{12} & A_{13} \\ A_{21} & A_{22} & A_{23} \\ A_{31} & A_{32} & A_{33} \end{smallmatrix} \Bigr)^{(2)} \\ \Bigl( \begin{smallmatrix} \rho J_1 \\ \rho J_2 \\ \rho J_3 \end{smallmatrix} \Bigr)^{(1)} \\ \Bigl( \begin{smallmatrix} \rho J_1 \\ \rho J_2 \\ \rho J_3 \end{smallmatrix} \Bigr)^{(2)}
\end{pmatrix}^{**K} &=  \quad \begin{pmatrix} 
z^{(1)} \rho^{(1)} \\ z^{(2)} \rho^{(2)} \\ \Bigl( \begin{smallmatrix} \rho u_1 \\ \rho u_2 \\ \rho u_3 \end{smallmatrix}\Bigr) \\ z^{(1)} \\ z^{(2)} \rho^{(2)} \lambda \\ \rho E  \\ \Bigl( \begin{smallmatrix} A_{11} & A_{12} & A_{13} \\ A_{21} & A_{22} & A_{23} \\ A_{31} & A_{32} & A_{33} \end{smallmatrix} \Bigr)^{(1)} \\ \Bigl( \begin{smallmatrix} A_{11} & A_{12} & A_{13} \\ A_{21} & A_{22} & A_{23} \\ A_{31} & A_{32} & A_{33} \end{smallmatrix} \Bigr)^{(2)} \\ \Bigl( \begin{smallmatrix} \rho J_1 \\ \rho J_2 \\ \rho J_3 \end{smallmatrix} \Bigr)^{(1)} \\ \Bigl( \begin{smallmatrix} \rho J_1 \\ \rho J_2 \\ \rho J_3 \end{smallmatrix} \Bigr)^{(2)}
\end{pmatrix}^{*K} + \frac{1}{S_t^K - S_c} \begin{pmatrix} 0\\ 0 \\ \Bigl( \begin{smallmatrix} 0\\ \sigma_{21}^* - \sigma_{21}^{**} \\ \sigma_{31}^* - \sigma_{31}^{**} \end{smallmatrix} \Bigr) \\ 0 \\ 0 \\ \sigma_{21}^* u_2^* + \sigma_{31}^* u_3^* - \sigma_{21}^{**} u_2^{**} - \sigma_{31}^{**} u_3^{**} \\ \Bigl( \begin{smallmatrix} A_{12} & 0 & 0 \\ A_{22} & 0 & 0 \\ A_{32} & 0 & 0 \end{smallmatrix} \Bigr)^{*(1)} (u_2^{**} - u_2^*) + \Bigl( \begin{smallmatrix} A_{13} & 0 & 0 \\ A_{23} & 0 & 0 \\ A_{33} & 0 & 0 \end{smallmatrix} \Bigr)^{*(1)} (u_3^{**} - u_3^*) \\ \Bigl( \begin{smallmatrix} A_{12} & 0 & 0 \\ A_{22} & 0 & 0 \\ A_{32} & 0 & 0 \end{smallmatrix} \Bigr)^{*(2)} (u_2^{**} - u_2^*) + \Bigl( \begin{smallmatrix} A_{13} & 0 & 0 \\ A_{23} & 0 & 0 \\ A_{33} & 0 & 0 \end{smallmatrix} \Bigr)^{*(2)} (u_3^{**} - u_3^*)\\ \Bigl( \begin{smallmatrix} 0 \\ 0 \\ 0 \end{smallmatrix} \Bigr)^{(1)} \\ \Bigl( \begin{smallmatrix} 0 \\ 0 \\ 0 \end{smallmatrix} \Bigr)^{(2)} \end{pmatrix}^K
\end{align}
From the transverse components of the momentum Rankine-Hugoniot equation, we have
\begin{align}
{\sigma_{j1}^{**}}_L = {\sigma_{j1}^{**}}_R = \frac{{\rho^*}_L {\rho^*}_L \left((u_j)_R - (u_j)_L\right) \left(S_t^L - S_c\right) \left(S_t^R - S_c\right) + {\rho^*}_L \left(S_t^L - S_c\right) (\sigma_{j1})_R - {\rho^*}_R \left(S_t^R - S_c\right) (\sigma_{j1})_L}{{\rho^*}_L \left(S_t^L - S_c\right) - {\rho^*}_R \left(S_t^R - S_c\right)},
\end{align}
for $j \neq 1$. Within the two middle states, we also have
\begin{subequations}
\begin{align}
{p^{**K}} - {\sigma_{11}^{**K}} &= {p^*K} - {\sigma_{11}^*K},\\*
T^{**K} &= T^{*K},\\*
q_1^{**K} &= q_1^{*K}. 
\end{align}
\end{subequations}
The approximate HLLD flux between states $\mathbf{U}_L$ and $\mathbf{U}_R$ is given by 
\begin{equation}
\mathbf{F}_{HLLD}(\mathbf{U}_L, \mathbf{U}_R) = 
\begin{cases}
\mathbf{F}(\mathbf{U}_L) &\text{if} \quad 0 \leq S_l^L\\
\mathbf{F}(\mathbf{U}_L) + S_l^L (\mathbf{U}_L^* - \mathbf{U}_L) &\text{if} \quad S_l^L < 0 \leq S_t^L\\
\mathbf{F}(\mathbf{U}_L) + S_l^L (\mathbf{U}_L^* - \mathbf{U}_L) + S_t^L (\mathbf{U}_L^{**} - \mathbf{U}_L^{*}) & \text{if} \quad S_t^L < 0 \leq S_c\\
\mathbf{F}(\mathbf{U}_R) + S_l^R (\mathbf{U}_R^* - \mathbf{U}_R) + S_t^R (\mathbf{U}_R^{**} - \mathbf{U}_R^{*}) & \text{if} \quad S_c < 0 < S_t^R\\
\mathbf{F}(\mathbf{U}_R) + S_l^R (\mathbf{U}_R^* - \mathbf{U}_R) & \text{if} \quad S_t^R \leq 0 < S_l^R\\
\mathbf{F}(\mathbf{U}_R) & \text{if} \quad S_l^R \leq 0
\end{cases}
\end{equation}
The contact wave speed is calculated using:
\begin{align}
S_c = \frac{\rho_R (u_1)_R \left((u_1)_R - S_l^R\right) - \rho_L (u_1)_L \left((u_1)_L - S_l^L\right) + p_R - (\sigma_{11})_R - p_L + (\sigma_{11})_L}{\rho_R \left((u_1)_R - S_l^R\right) - \rho_L \left((u_1)_L - S_l^L\right)},
\end{align}
whereas the normal thermal impulse components in the star states are, for $l = 1, 2$
\begin{align}
	J_c^{(l)} = \frac{\rho_R (J_1^{(l)})_R \left((u_1)_R - S_l^R\right) - \rho_L (J_1^{(1)})_L \left((u_1)_L - S_l^L\right) + T^{(l)}_R - T^{(l)}_L}{\rho_R \left((u_1)_R - S_l^R\right) - \rho_L \left((u_1)_L - S_l^L\right)}. 
\end{align}
The remaining task is to approximate the speeds of the four outermost waves, $S_l^L, S_t^L, S_t^R, S_l^R$. 
Following Dumbser and Balsara \cite{dumbserHLLEM}, the longitudinal wave speeds are taken to be 
\begin{align}
&S_l^L = \text{min}\left(\mathbf{\Lambda}(\mathbf{U}_L), \mathbf{\Lambda}(\bar{\mathbf{U}})\right), &S_l^R = \text{max}\left(\mathbf{\Lambda}(\mathbf{U}_R), \mathbf{\Lambda}(\bar{\mathbf{U}})\right),
\end{align}
where the intermediate state $\bar{\mathbf{U}}$ is an arithmetic average, $\bar{\mathbf{U}} = \frac{1}{2} (\mathbf{U}_L + \mathbf{U}_R)$ and $\mathbf{\Lambda}(\mathbf{U})$ denotes the diagonal matrix of eigenvalues of the system's characteristic matrix. The transverse wave speeds are chosen following the work of Ortega et al.\ \cite{ortega} 
\begin{align} 
&S_t^L = u_1^* - c_s(\mathbf{U}^*_L), &S_t^R = u_1^* + c_s(\mathbf{U}^*_R),
\end{align}
where, in our multi-phase formulation, the shear sound speed of the global mixture, $c_s(\mathbf{U})$ is mass weighted. 
Finally, the solver is extended to second order of accuracy using a typical MUSCL-Hancock reconstruction scheme \cite{toro}. 

\subsection{ODE solver}
The source ODE system \eqref{eq:source} can be written as 
\begin{subequations}
\begin{align}
\frac{d \lambda}{d t} &= \mathcal{K},\label{eq:lambdaODE}\\*
\frac{d A^{(l)}}{dt} &= - \frac{3}{\tau_1^{(l)}} |A^{(l)}|^\frac{5}{3} A^{(l)} \text{dev}({G}^{(l)}),\label{eq:AODE}\\*
\frac{d \mathbf{J}^{(l)}}{dt} &= - \frac{1}{\tau_2^{(l)}} \frac{T^{(l)}}{T_0^{(l)}} \frac{\rho_0^{(l)}}{\rho} \mathbf{J}^{(l)},\label{eq:JODE} 
\end{align}
\end{subequations}
for $l = 1, 2$. These are governed by different timescales, depending on the relaxation times for the distortion tensors and thermal impulse vectors. For example, if $\tau_1^{(l)}$ or $\tau_2^{(l)}$ is very small, then the corresponding ODE is stiff, which would require a small time-step in its solution. 

We therefore split these ODEs using Strang's splitting, as before. This then enables us to use an appropriate solver for each ODE. Equation \eqref{eq:strang} is therefore replaced by
\begin{align}
\mathbf{U}^{n+1} = \mathcal{R}^{\frac{\Delta t}{2}} \mathcal{D}^{\frac{\Delta t}{2}} \mathcal{J}^{\frac{\Delta t}{2}} \mathcal{H}^{\Delta t} \mathcal{J}^{\frac{\Delta t}{2}} \mathcal{D}^{\frac{\Delta t}{2}} \mathcal{L}^{\frac{\Delta t}{2}} \mathbf{U}^n,
\end{align}
where $\mathcal{R}^{\Delta t}, \mathcal{D}^{\Delta t}, \mathcal{J}^{\Delta t}$ are the operators evolving equations \eqref{eq:lambdaODE}, \eqref{eq:AODE}, \eqref{eq:JODE} by $\Delta t$, respectively. The reactive ODE is solved using an explicit fourth order Runge-Kutta method. The distortion and thermal impulse equations \eqref{eq:AODE}, \eqref{eq:JODE} are solved using the semi-analytic ODE solver of Jackson \cite{jackson, jackson19}. 

\section{Validation and evaluation}
\label{sect:validation}
\begin{table}[!tb]
	\centering
	\caption{Test suite employed for validation and evaluation purposes, with corresponding properties and effective reduced model. Tests marked with a hyphen under the section column, \#, have been performed as validation but omitted from this paper for brevity. Material types `inviscid' and `viscous' refer to fluid materials, whereas `elastic' and `elastoplastic' refer to solid materials. }
	\label{table:testsuite}
	\scriptsize
	\begin{tabular}{l l l l l l l}
		\textbf{\#} & \textbf{Problem} & \textbf{Reduced model} & \textbf{Components} & \textbf{Reaction} & \makecell[l]{\textbf{Material}\\\textbf{type}} & \makecell[l]{\textbf{Heat}\\ \textbf{Conduction}}\\ 
		\hline \\[-4pt]
		- & TNT products shock tube \cite{shyue} & Euler equations & 1 & No & Inviscid & No \\
		- & Copper 7-wave \cite{barton09} &  Miller \& Colella \cite{millercolella} & 1 & No & Elastic & No\\
		- & Copper piston \cite{peshkov} & GPR \cite{dumbser} & 1 & No & Elastoplastic & No\\
		- & Stokes' first problem \cite{dumbser} & GPR \cite{dumbser} & 1 & No & Viscous & No\\
		- & Viscous shock \cite{dumbser} & GPR \cite{dumbser} & 1 & No & Viscous & Yes\\
		- & Heat conduction in a gas \cite{dumbser} & GPR \cite{dumbser} & 1 & No & Viscous & Yes\\
		\hline \\[-4pt]
		- & Copper - TNT shock tube \cite{saurel99} & Allaire et al.\ \cite{allaire} & 2 & No & Inviscid & No\\
		- & PBX-9404 - copper shock tube \cite{barton11} & Model I (Fig. \ref{fig:limiting}) & 2 & No & \makecell[l]{Inviscid, \\Elastic} & No \\
		- & Aluminium-copper `stick' test \cite{barton19solidfluid} & Model II (Fig. \ref{fig:limiting}) & 2 & No & Elastic & No\\
		\ref{sect:LX-17} & Shock induced ignition in LX-17 \cite{kapila} & Banks et al.\ \cite{banks08} & 2 & Yes & Inviscid & No\\
		\ref{sect:C4} & ZND detonation propagation in C-4 \cite{michael18} & Banks et al.\ \cite{banks08} & 2 & Yes & Inviscid & No\\
		\ref{sect:viscousdetonation} & Viscous Shock-Induced Detonation \cite{hidalgo2011ader} & Augmented GPR \cite{jacksonthesis} & 2 & Yes & Viscous & Yes\\
		\ref{sect:heatdetonation} & Viscous Heating-Induced Detonation \cite{clarke2} & Augmented GPR \cite{jacksonthesis} & 2 & Yes & Viscous & Yes\\
		\hline\\[-4pt]
		\ref{sect:hydroconfined} & Detonation in LX-17 confined by fluid \cite{hybrid} & MiNi16 \cite{hybrid} & 3 & Yes & Inviscid & No\\
		\ref{sect:elastoplasticconfined} & Detonation in LX-17 confined by copper & Full model & 3 & Yes & \makecell[l]{Inviscid,\\ Elastoplastic} & No\\
		\ref{sect:heatdetonation_confined} & \makecell[l]{Viscous Heating-Induced Detonation \\confined by water} & Full model & 3 & Yes & \makecell[l]{Viscous,\\Inviscid} & Yes
	\end{tabular}
\end{table}
\begin{table}
	\centering
	\caption{Equation of state parameters for materials considered in the validation and evaluation tests. These are scaled with respect to different reference values and are hence non-dimensional (see section describing the corresponding test).}
	\label{table:EOSparams}
	\footnotesize
	\begin{tabular}{l c c c c c c c}
		\hline\\[-7.5pt]
		\textbf{JWL parameters} & $\boldsymbol{\rho_0}$ & $\boldsymbol{\Gamma_0}$ & $\boldsymbol{\mathcal{A}}$ & $\boldsymbol{\mathcal{B}}$ & $\boldsymbol{\mathcal{R}_1}$ & $\boldsymbol{\mathcal{R}_2}$ & $\boldsymbol{C_v}$\\
		LX-17 reactants & 1 & 0.8939 & 692.5 & -0.04478 & 11.3 & 1.13 & 0.006596\\
		LX-17 products & 1 & 0.5 & 13.18 & 0.5677 & 6.2 & 2.2 & 0.002652\\
		C-4 reactants & 1 & 0.8938 & 751.7874 & -0.04861 & 11.3 & 1.13 & 0.00716\\
		C-4 products & 1 & 0.25 & 5.8915 & 0.1251 & 4.5 & 1.4 & 0.002879\\
		Hydrodynamic confiner & 1 & 0.8 & 100.4 & -0.04 & 10 & 1.5 & 0.006596\\
		\hline\\[-7.5pt]
		\textbf{Shock Mie-Gr\"uneisen parameters} & $\boldsymbol{\rho_0}$ & $\boldsymbol{\Gamma_0}$ & $\boldsymbol{c_0}$ & $\boldsymbol{s}$ & $\boldsymbol{N}$ & &\\
		Copper & 4.6877 & 2 & 0.513 & 1.48 & 0 & &\\
		\hline\\[-7.5pt]
		\textbf{Ideal gas parameters} & $\boldsymbol{\rho_0}$ & $\boldsymbol{\gamma}$ & $\boldsymbol{C_v}$ & $\mathbf{T}_0$ & & &\\
		Viscous gas & 1 & 1.4 & 2.5 & 1 & & &\\
		\hline\\[-7.5pt]
		\textbf{Stiffened gas parameters} & $\boldsymbol{\rho_0}$ & $\boldsymbol{\gamma}$ & $\boldsymbol{p_\infty}$ & & & &\\
		Water & 849.91 & 4.4 & 5921.54 & & & &\\
		\hline
	\end{tabular}
\end{table}
In this section we validate the new formulation against various tests from the literature. 
As discussed earlier, the presented formulation reduces to a number of limiting models. Therefore, tests that are effectively solved using the reduced models can be used as validation; see Figure \ref{fig:limiting} for examples of possible applications. An extensive validation of the limiting models has been performed, using a broad range of  test cases from the literature. However here we only present a selection of validation and evaluation tests for the sake of brevity. The tests employed for validation and evaluation (including omissions), along with their properties and corresponding reduced models, are presented in Table \ref{table:testsuite}.

The materials considered range in mechanical properties from inviscid to elastic and elastoplastic. As a result, a number of different equations of state are used to describe each material. These will be introduced in the sections describing the corresponding test. Unless stated otherwise, heat conduction is neglected, therefore we set $c_t^{(l)} = 0$ and initialise the thermal impulse vectors as $\mathbf{J}^{(l)} = \mathbf{0}$. 

As for visualisation, in problems with material interfaces we use two distinct colours for the materials on either side of the interface. This is achieved by having a linear dependence of colour on the volume fraction, $z^{(1)}$. As a result, the position and width of the numerically diffused interface are both clearly visible in the plots. 

Transmissive boundary conditions are assumed at both boundaries, except where indicated otherwise. 

\subsection{LX-17 shock initiation problem}
\label{sect:LX-17}
This work is particularly focused on the modelling of elastoplastically confined combustion. It is therefore a vital step to validate the formulation in problems involving the initiation and propagation of detonation waves. Doing so, we are interested in making sure that the ignition mechanism, as well as the numerically calculated properties of the ZND detonation match with experiments. For these applications, the full formulation reduces to the reactive fluid-mixture model of Banks et al.\ \cite{banks08}.  

In this test, a slab of the plastic-bonded explosive LX-17, is slammed impulsively into a stationary wall, causing a shock wave to form and propagate in the explosive. This in turn leads to the initiation of the material. The stationary wall is simulated in the form of reflective left boundary conditions. Transmissive conditions are used for the right boundary. Both reactants and products are governed by the JWL equation of state, which is commonly used in detonation modelling, 
\begin{equation}
\begin{alignedat}{2}
p_{ref} (\rho) &= \mathcal{A} \exp{\Big( - \mathcal{R}_1 \frac{\rho_0}{\rho} \Big)} + \mathcal{B} \exp{\Big( -\mathcal{R}_2 \frac{\rho_0}{\rho} \Big)}, \qquad \qquad \quad T_{ref}(\rho) &&= 0,\\
e_{ref} (\rho) &= \frac{\mathcal{A}}{\mathcal{R}_1 \rho_0}\exp{\Big( - \mathcal{R}_1 \frac{\rho_0}{\rho} \Big)} + \frac{\mathcal{B}}{\mathcal{R}_2 \rho_0}  \exp{\Big( -\mathcal{R}_2 \frac{\rho_0}{\rho} \Big)}, \qquad \quad \Gamma(\rho) &&= \Gamma_0.
\end{alignedat}
\label{eq:JWL}
\end{equation}
The equation of state parameters for LX-17 reactants and products are given in Table \ref{table:EOSparams}. The heat of combustion is $\mathcal{Q} = 0.06141$. LX-17 is modelled as an inviscid fluid, therefore we set $\tau_1 = 0$ and $c_s = 0$.  All parameters have been scaled with respect to the LX-17 reference state \cite{kapila}, 
\begin{equation}
\begin{alignedat}{3}
	&\tilde{\rho} = \rho_0 = 1905\, \si{\kilogram \per \cubic \meter},\qquad  &&\tilde{u} = D_{CJ} = 7679.9473\, \si{\meter \per \second}, \qquad &&\tilde{p} = D_{CJ}^2 \rho_0 = 1.1235993\times 10^{11}\, \si{\pascal}, \\*
	&\tilde{t} = 10^{-6}\, \si{\second}, && \tilde{l} = D_{CJ} \tilde{t} = 7.6799473\, \si{\milli \meter}, \qquad && \tilde{T} = 298\, \si{\kelvin},\\*
	&\tilde{C_v} = \tilde{p}/\tilde{T},  && && 
\end{alignedat}\label{eq:LX-refstate}
\end{equation}
and are hence non-dimensional. The Ignition and Growth reaction rate law from \cite{ignitiongrowth} is used, 
\begin{align}
\begin{split}
	\frac{d \phi}{dt} = - \mathcal{K} = I (1 &- \phi)^b (\rho - 1 - a)^x H(\phi_{IG_{max}} - \phi)\\
	&+ G_1 (1 - \phi)^c \phi^d p^y H(\phi_{{G_1}_{max}} - \phi)\\
	&+ G_2 (1 - \phi)^e \phi^g p^z H(\phi - \phi_{{G_2}_{max}}),
\end{split}\label{eq:IgnitionGrowth}
\end{align}
with the following non-dimensional parameters, 
\begin{align}
	&I = 4 \times 10^6, && b = 0.667, && a = 0.22, && x = 7, &&\phi_{IG_{max}} = 0.02,\nonumber\\*
	& G_1 = 6383.3241, && c = 0.667, && d = 1, && y = 3, && \phi_{{G_1}_{max}} = 0.8,\label{eq:IG-LX-17}\\*
	&G_2 = 33.708, &&e = 0.667, && g = 0.667, && z = 1, &&\phi_{{G_2}_{max}} = 0.8.\nonumber
\end{align}
The test's domain is $[0, 5]$ and the left boundary is modelled as a reflective wall. Initially, the explosive is moving towards the wall, with the following non-dimensional state variables
\begin{align*}
\big(\rho^{(1)},\,\, \rho^{(\alpha)},\,\, \rho^{(\beta)},\,\, \mathbf{u},\,\, z^{(1)},\,\, \lambda,\,\, p,\,\, A^{(1)},\,\, A^{(2)},\,\, \mathbf{J}^{(1)},\,\, \mathbf{J}^{(2)}\big) =
\big(1,\,\, 1,\,\, 1,\,\, \left(-0.14 \quad 0 \quad 0 \right),\,\, 10^{-6},\,\, 1,\,\, 0.001,\,\, I_3,\,\, I_3,\,\, \mathbf{0},\,\, \mathbf{0}\big). 
\end{align*}

\begin{figure}
	\centering
	\includegraphics[width=0.8\linewidth]{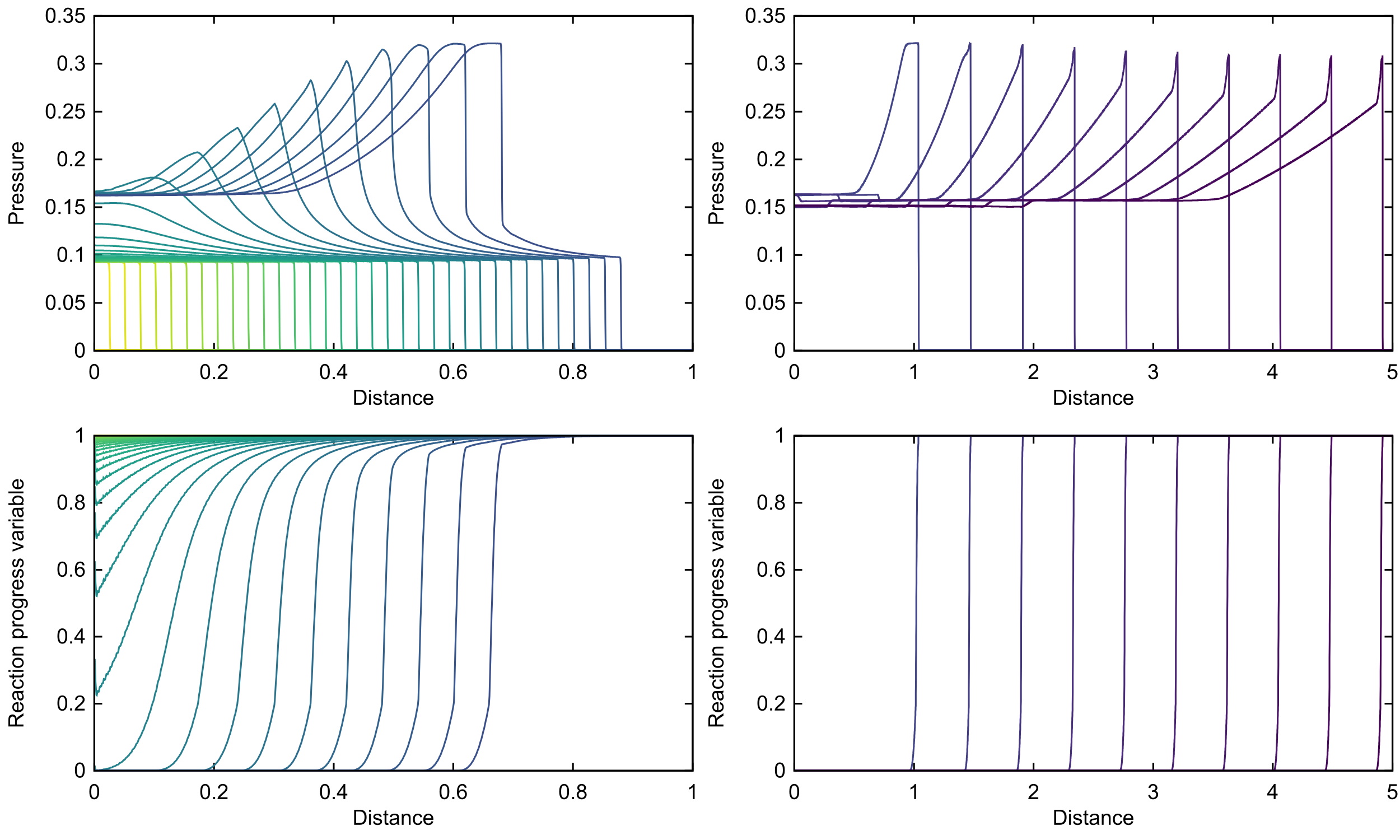}
	\caption{Superimposed plots for non-dimensional pressure (top) and reaction progress variable $\lambda$ (bottom) for the LX-17 shock initiation problem at times $t = 0\, (0.05)\, 1.7$ (left) and $t = 2\, (0.5)\, 6.5$ (right). }
	\label{fig:LX-shock}
\end{figure}

The results are shown in Figure \ref{fig:LX-shock}, where plots of pressure and $\lambda$ are superimposed at different stages of the reaction process. Numerical solutions are calculated using a non-dimensional step size of $\Delta x = 0.000625$. The two left plots present profiles between $t = 0$ and $1.7$. During this period, an initial shock wave forms from the impact of the material on the wall. Density increases, reaching a threshold of $1 + a = 1.22$, which activates the ignition term in the reaction rate law. The energy released from the reaction process causes the pressure near the wall to increase, forming a pulse that propagates inwards. This pulse is amplified, steepened and eventually forms a second shock wave, behind the leading shock. The stronger, first growth term is active until $\lambda$ reaches a value of $0.2$. At this point, it is switched off, triggering the weaker, second growth term. Also, the reaction rate immediately drops, as suggested by the discontinuous change in the gradient of $\lambda$. Finally, in the last stages of evolution (two right plots), the secondary shock overtakes the lead shock and eventually the structure gradually approaches the CJ state from above, with $p_{CJ} = 0.240297$ ($0.269998 \times 10^{11} \si{\pascal}$ in dimensional terms). The von Neumann spike also approaches the von Neumann pressure $p_{VN} = 0.309929$ ($0.348236 \times 10^{11} \si{\pascal}$). 

Despite using a different closure condition (constant density ratio in place of temperature equilibrium between reactants and products), great agreement is achieved with the corresponding results of Kapila et al.\ \cite{kapila}. 

\subsection{C-4 ZND propagation problem}
\label{sect:C4}
\begin{figure}[!tb]
	\centering
	\captionof{table}{Initial conditions (non-dimensional) for the C-4 ZND propagation problem}
	\label{table:C4}
	\begingroup
	\setlength{\tabcolsep}{11.7pt} 
	\renewcommand{\arraystretch}{1.1} 
	\scriptsize
	\begin{tabular}{l l c c c c c c c}
		\hline\\[-7pt]
		Material & Region & $\rho^{(1)}, \rho^{(\alpha)}, \rho^{(\beta)}$ & $\mathbf{u}$ & $z^{(1)}$ & $\lambda$ & $p$ & $A^{(1)}, A^{(2)}$ & $\mathbf{J}^{(1)}, \mathbf{J}^{(2)}$\\[2pt] \hline \\[-6pt]
		C-4 (Booster) & $\left[0, 0.05\right]$ & 0.9931 & $\mathbf{0}$ & $10^{-6}$ & 0 & $0.386473$ & $I_3$ & $\mathbf{0}$\\
		C-4 & $\left(0.05, 8\right]$ & 0.9931 & $\mathbf{0}$ & $10^{-6}$ & 1 & $9.6618 \times 10^{-7}$ & $I_3$ & $\mathbf{0}$\\
		\hline
	\end{tabular}
	\endgroup
	
	\includegraphics[width=0.65\linewidth]{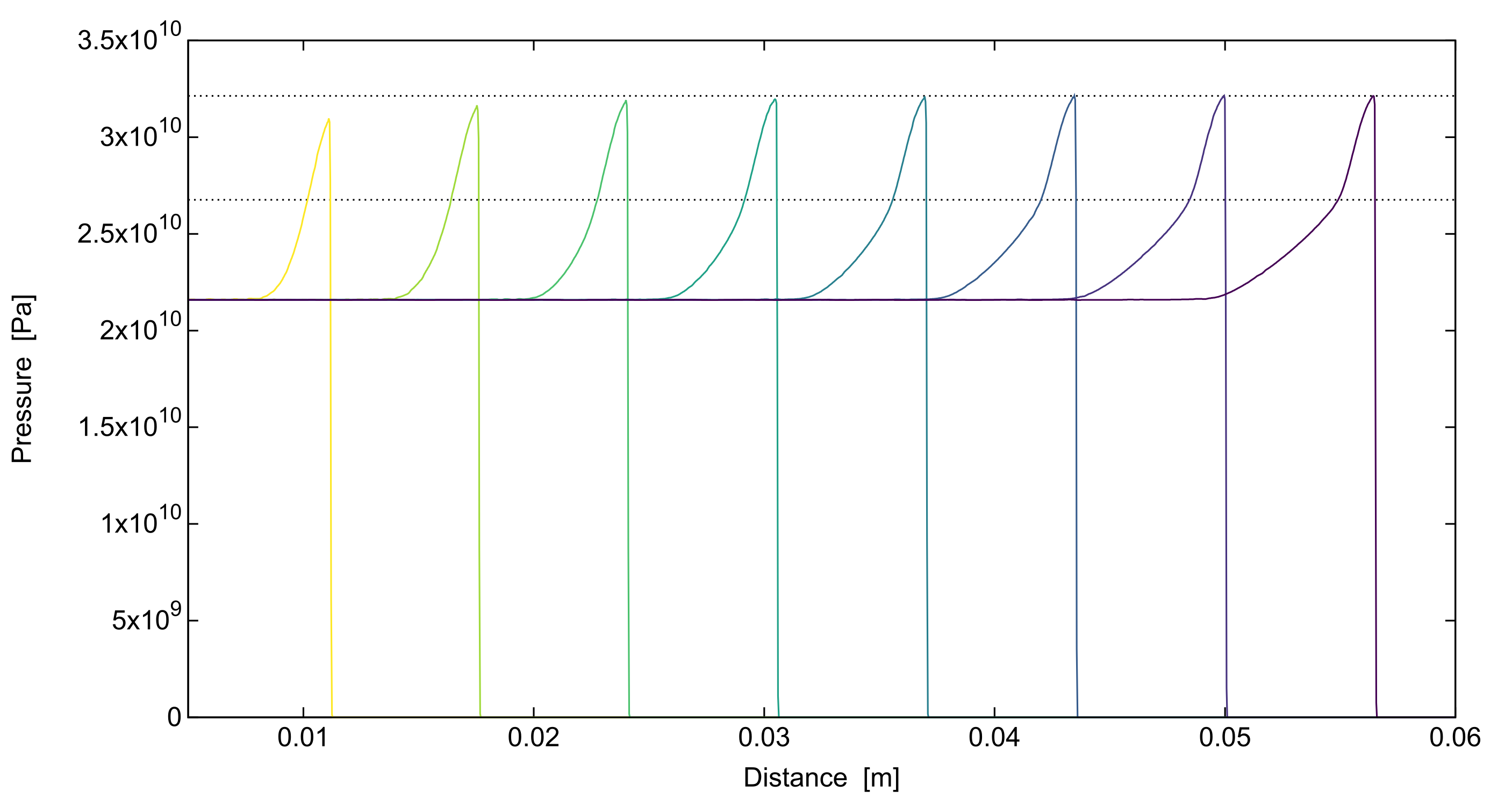}
	\captionof{figure}{Superimposed pressure profiles of a propagating detonation in C-4, approaching a steady state. The horizontal lines denote the von Neumann and CJ pressures. }
	\label{fig:C4_znd}
\end{figure}
In this test, we verify that the classical, steady ZND detonation structure is achieved under appropriate initial conditions for the explosive C-4. As in the previous test, we effectively solve the reactive fluid-mixture model of Banks et al.\ \cite{banks08}. The equations of state for both the reactants and the products of C-4 follow a JWL form \eqref{eq:JWL}, with parameters taken from \cite{C4params} and summarised in Table \ref{table:EOSparams}. The heat of combustion is $\mathcal{Q} = 0.0870$. C-4 is modelled as an inviscid fluid, therefore we set $\tau_1 = 0$ and $c_s = 0$. The Ignition and Growth reaction rate law \eqref{eq:IgnitionGrowth} is used, with non-dimensional parameters 
\begin{align}
&I = 4\times 10^6, && b = 0.667, && a = 0.0367, && x = 7, &&\phi_{IG_{max}} = 0.022,\nonumber\\*
& G_1 = 149.97, && c = 0.667, && d = 0.33, && y = 2, && \phi_{{G_1}_{max}} = 1,\\*
&G_2 = 0, &&e = 0.667, && g = 0.667, && z = 3, && \phi_{{G_2}_{max}} = 0.\nonumber 
\end{align}
Both the equation of state parameters and the reaction parameters have been non-dimensionalised with respect to the following reference variables, 
\begin{equation}
\begin{alignedat}{3}
&\tilde{\rho} = \rho_0 = 1601\, \si{\kilogram \per \cubic \meter}, &&\tilde{u} = D_{CJ} = 8040\, \si{\meter \per \second}, \qquad &&\tilde{p} = D_{CJ}^2 \rho_0 = 1.035\times 10^{11}\, \si{\pascal}, \\*
&\tilde{t} = 10^{-6}\, \si{\second}, && \tilde{l} = D_{CJ} \tilde{t} = 8.04\, \si{\milli \meter}, \qquad && \tilde{T} = 298\, \si{\kelvin},\\*
&\tilde{C_v} = \tilde{p}/\tilde{T} = 3.4729 \times 10^8\, \si{\pascal \per \kelvin}. \qquad && && 
\end{alignedat}
\end{equation}
We initiate the explosive by means of a booster, i.e.\ a slab of explosive raised at $40\, \si{\giga \pascal}$, see Table \ref{table:C4} for the initial conditions used. A non-dimensional cell size of $\Delta x = 0.004167$ is used, which in dimensional terms corresponds to $\Delta x = 33.5\, \si{\micro \meter}$. 

The solution is expected to converge to a steady state ZND detonation. Figure \ref{fig:C4_znd} depicts the detonation wave, in the form of dimensional pressure profiles at equal time steps. The fact that the last few pressure profiles are self-similar and separated by an equal distance implies steadiness. The computed converged CJ and von Neumann pressure values are $ 26.763\, \si{\giga \pascal}$ and $32.137\, \si{\giga \pascal}$, respectively, as illustrated by the dotted lines in Figure \ref{fig:C4_znd}. The computed CJ pressure value falls well within the range of values found in the literature \cite{neuber}: $27.5\, \si{\giga \pascal}, 24.91\, \si{\giga\pascal}, 22.55\, \si{\giga \pascal}, 25.09\, \si{\giga\pascal}$ and $22.36\,  \si{\giga \pascal}$. Our computed ZND properties are also close to those computed by Michael and Nikiforakis \cite{michael18} of CJ pressure at $27.5\, \si{\giga \pascal}$ and von Neumann pressure at $31.2\, \si{\giga \pascal}$, where a temperature equilibrium condition was imposed between reactants and products. 

\subsection{Viscous Shock-Induced Detonation}
\label{sect:viscousdetonation}
Having modelled detonations in inviscid explosives, it is worth considering materials exhibiting reactive, viscous and heat conducting behaviour. The full model reduces in this case to the GPR model, augmented with a reaction equation. For this specific problem, the same equation of state and parameters are used for reactants and products, therefore the reduced model corresponds to the reactive GPR model developed by Jackson \cite{jacksonthesis}. 

\begin{figure}[!tb]
	\centering
	\captionof{table}{Initial conditions (non-dimensional) for the viscous shock-induced detonation problem. The discontinuity position is $x_c = 0 / x_c = 0.5$ for the high/low viscosity case, respectively. }
	\label{table:viscousdetonation_ICs}
	\begingroup
	\setlength{\tabcolsep}{13pt} 
	\renewcommand{\arraystretch}{1.1} 
	\scriptsize
	\begin{tabular}{l c c c c c c c}
		\hline\\[-7pt]
		Region & $\rho^{(1)}, \rho^{(\alpha)}, \rho^{(\beta)}$ & $\mathbf{u}$ & $z^{(1)}$ & $\lambda$ &  $p$ & $A^{(1)}, A^{(2)}$ & $\mathbf{J}^{(1)}, \mathbf{J}^{(2)}$\\[2pt] \hline \\[-6pt]
		$x < x_c$ & 
		$1.4$ & $\mathbf{0}$ & $10^{-6}$ & 0 & 1 & $\sqrt[3]{1.4}I_3$ & $\mathbf{0}$\\[3pt]
		$x_c \leq x$ & $0.887565$ &  
		$\begin{pmatrix}
		-0.57735& 0& 0
		\end{pmatrix}^\intercal$ & $10^{-6}$ & 1 & 0.191709 & $\sqrt[3]{0.887565}I_3$ & $\mathbf{0}$\\
		\hline
	\end{tabular}
	\endgroup
	
	\vspace{10pt}
	\begingroup
	\begin{subfigure}[b]{\textwidth}
		\centering
		\includegraphics[width=\linewidth]{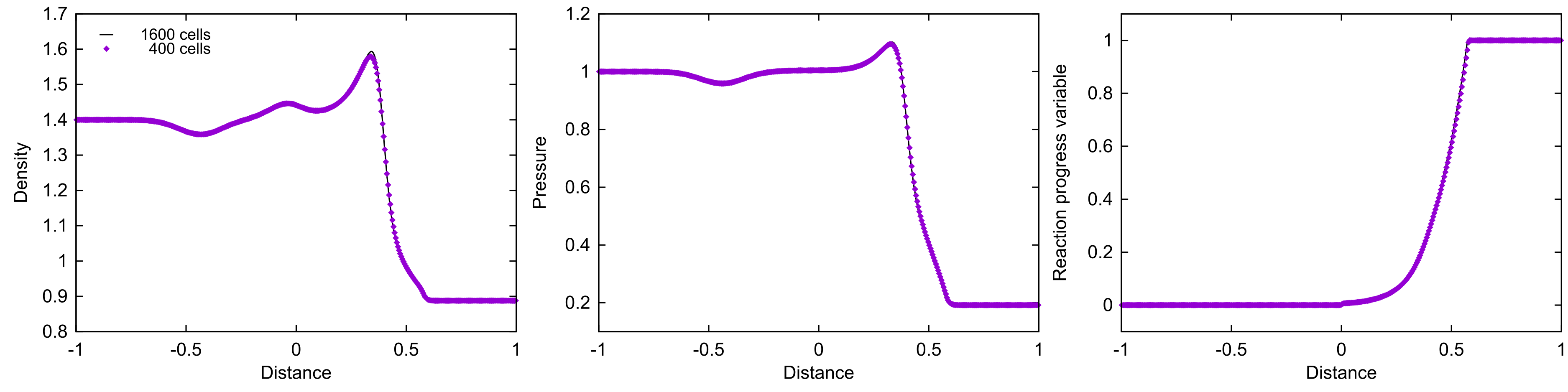}
	\end{subfigure}
	\begin{subfigure}[b]{\textwidth}
		\centering
		\includegraphics[width=\linewidth]{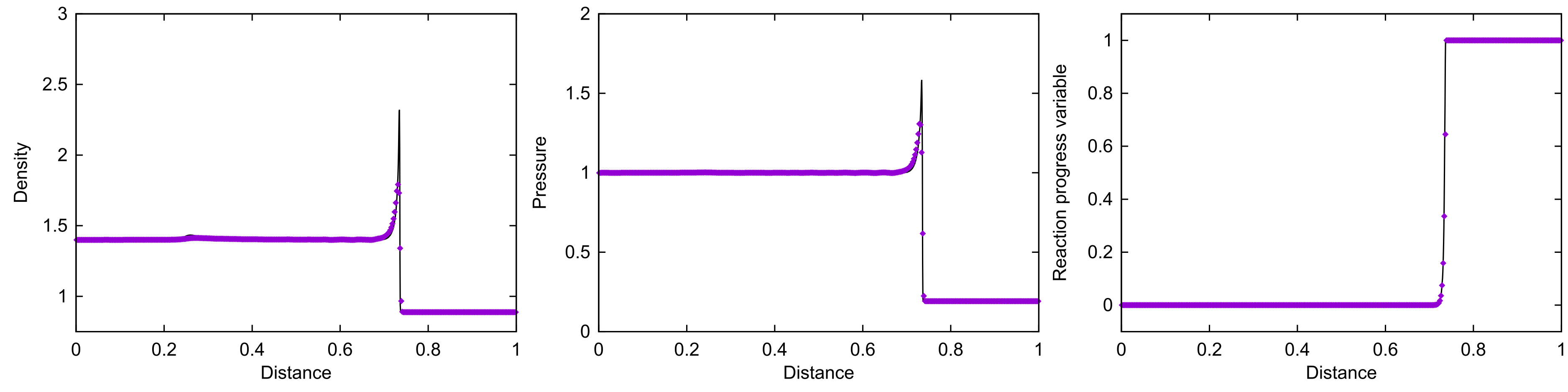}
	\end{subfigure}
	\captionof{figure}{Viscous shock-induced detonation problem. Numerical solutions for non-dimensional density, pressure and reactants concentration with 400 cells (points) plotted against numerical solution with 1600 cells (solid line) at $t = 0.5$. \textit{Top row: }High viscosity case with $\mu = 10^{-2}$ and $K_0 = 10$. \textit{Bottom row: } Low viscosity case with $\mu = 10^{-4}$ and $K_0 = 250$. }
	\label{fig:viscousdetonation}
	\endgroup
\end{figure}

The initial conditions for this test are taken from Hidalgo and Dumbser \cite{hidalgo2011ader} and correspond to an ideal Chapman-Jouguet wave travelling into a region of unburnt gas. Two different cases are considered; a more viscous case with lower reaction rate and a less viscous case with higher reaction rate. The non-dimensional initial conditions for both cases can be found in Table \ref{table:viscousdetonation_ICs}. The viscous gas is described by the ideal gas equation of state
\begin{equation}
\begin{alignedat}{2}
p_{ref}(\rho) &= 0, \qquad \qquad T_{ref}(\rho) &&= 0,\\
e_{ref}(\rho) &= 0, \qquad \qquad \quad \,\Gamma(\rho) &&= \gamma - 1,
\end{alignedat}
\label{eq:IG}
\end{equation}
with parameters given in Table \ref{table:EOSparams}. The heat of combustion is $\mathcal{Q} = 1$ and the Prandtl number is set to $\textit{Pr} = \gamma C_v \mu / \kappa = 0.75$. Hidalgo and Dumbser \cite{hidalgo2011ader} solved the compressible Navier-Stokes equations, therefore we have access to the viscosity coefficient and heat conductivity of the material of interest, but these alone are not sufficient to recover the parameters $c_t$ and $c_s$ of the GPR model. For this reason, we choose $c_t$ and $c_s$ so that our results correspond closely to those in \cite{hidalgo2011ader}. 

A simple, discrete ignition temperature kinetics model is considered, 
\begin{align}
\mathcal{K} = \begin{cases}
K_0 \lambda \quad &T \geq T_{ign},\\
0  & T < T_{ign},
\end{cases}
\end{align}
where $T_{ign}$ is the ignition temperature and $K_0$ is the inverse characteristic time scale of the chemical reaction. For this problem, we take $T_{ign} = 0.25$. 

For the high viscosity case, we set $\mu = 10^{-2},  c_s = 1$ and $c_t = 10$ for both the reactants and the products, and $K_0 = 10$. The problem's domain is $[-1, 1]$. The numerical results for density, pressure and mass fraction are depicted in the top row of Figure \ref{fig:viscousdetonation}. A great agreement is achieved both with the higher resolution results and with those of Hidalgo and Dumbser \cite{hidalgo2011ader}. As expected, having a high viscosity coefficient diffuses the combustion wave, resulting in a smooth ZND-like profile, with a rounded von Neumann spike.  It is important to mention that this test is initialised from the exact algebraic conditions for an ideal Chapman-Jouguet detonation wave. In these simulations, however, we take into account the effects of viscosity. This is the reason behind the leftward travelling waves in density and pressure. The same initialisation errors are present in \cite{hidalgo2011ader}. 

In the low viscosity test case, we take $\mu = 10^{-4}, c_s = 1$ and $c_t = 1$ for the reactants and products, and a higher reaction rate of $K_0 = 250$. The numerical results for density, pressure and mass fraction are depicted in the bottom row of Figure \ref{fig:viscousdetonation}, and a good agreement with the results of \cite{hidalgo2011ader, jacksonthesis} is obtained. As opposed to the high viscosity case, we can now clearly observe the ZND detonation structure, similarly to the inviscid problems \ref{sect:LX-17}, \ref{sect:C4}. Furthermore, the spatial resolution is now significant in resolving the sharp von Neumann spike. 

\subsection{Viscous Heating-Induced Detonation}
\label{sect:heatdetonation}
\begin{figure}[!tb]
	\centering
	\includegraphics[width=0.8\linewidth]{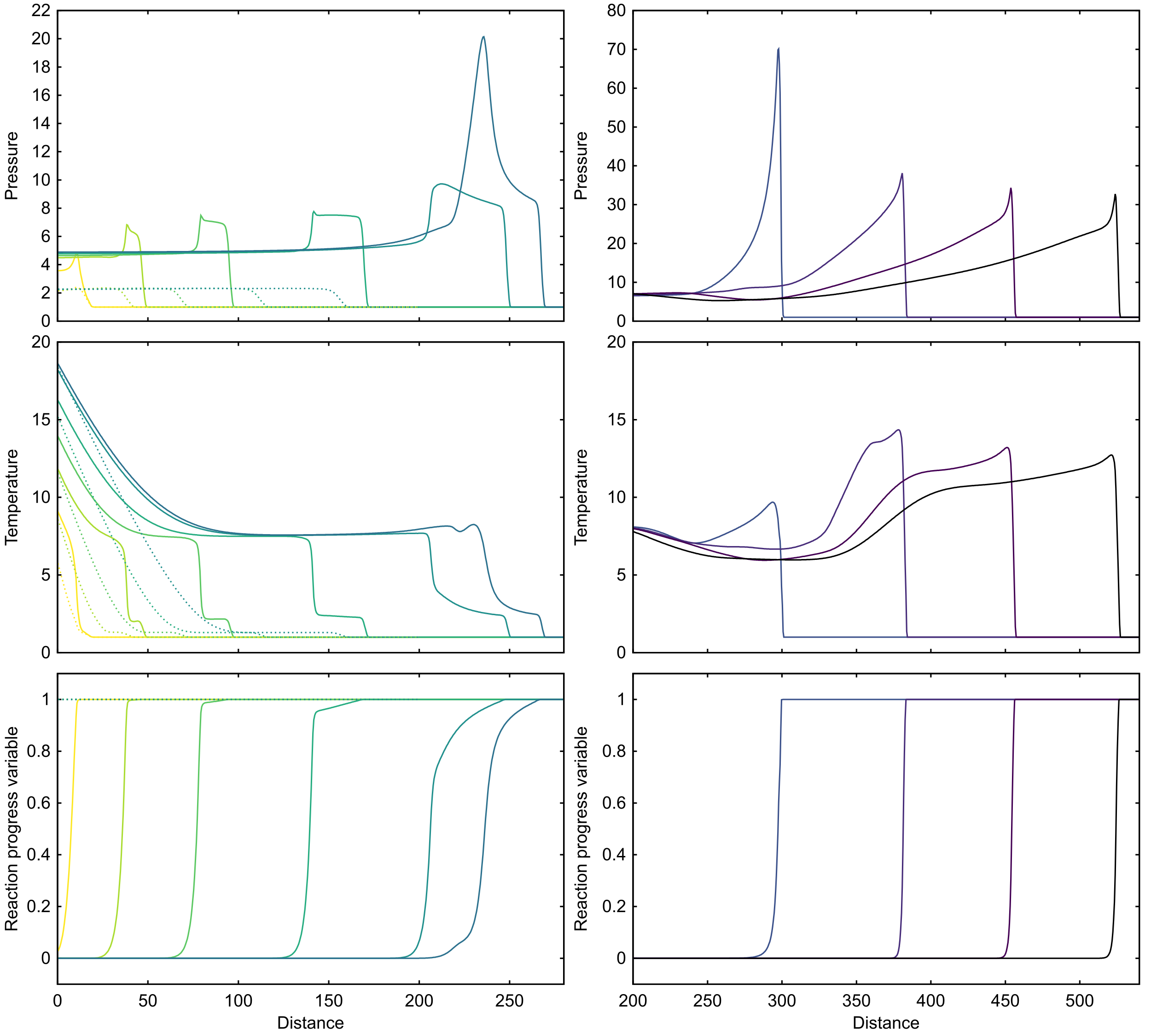}
	\caption{Profiles of dimensionless pressure, temperature and reactants concentration for the heating-induced detonation problem. Solid line plots correspond to the reactive case, at $t = 8.5, 21, 38, 63, 88.5, 94.5$ (left) and $t = 104, 114, 124, 134$ (right). Dotted line plots correspond to the inert case, at $t = 8.5, 21, 38, 63, 88.5$. }
	\label{fig:heatdetonation}
\end{figure}
So far, we have considered inviscid as well as viscous detonations, initiated by either a shock wave or a booster. Next, we will consider the heat-induced initiation of a detonation wave by the deposition of thermal energy at the boundary, as studied by Clarke et al.\ \cite{clarke2}. For this case, the model reduces to the GPR model augmented with a reaction equation of Jackson \cite{jacksonthesis}, as in the viscous, shock-induced detonation problem \ref{sect:viscousdetonation}. The purpose of this test is to demonstrate the significance of the inclusion of heat conduction in our model, when it comes to the initiation of a detonation wave. 

For this test, we scale all variables and parameters to render them non-dimensional. The reference state is given by
\begin{equation}
\begin{alignedat}{4}
&\tilde{\rho} = 1.1766 \,\si{\kilogram \per \cubic \meter}, \qquad \quad && \tilde{p} = 101325 \,\si{\pascal}, \qquad \quad &&\tilde{T} = 300 \,\si{\kelvin},  \qquad \quad &&\tilde{x} = \tilde{\mu_0} \tilde{c} / \tilde{p} \gamma,\\*
& \tilde{u} = \sqrt{\tilde{p}/\tilde{\rho}}, &&\tilde{J} = \tilde{T}/\tilde{\rho} \tilde{u}, &&\tilde{C_v} = \tilde{p}/\tilde{\rho} \tilde{T}, &&\tilde{\mu} = \tilde{\rho} \tilde{x} \tilde{u},
\end{alignedat}
\label{eq:heatdetonationrefstate}
\end{equation}
where $\tilde{c} = \sqrt{\gamma \tilde{p}/\tilde{\rho}}$ is the sound speed of the reference state and $\tilde{\mu_0} = 1.98 \times 10^{-5} \,\si{\pascal \second}$ is the coefficient of viscosity. 
Initially, the domain is occupied by an ideal gas at ambient conditions: 
\begin{align*}
\left(\rho^{(1)},\, \rho^{(\alpha)},\, \rho^{(\beta)},\, \mathbf{u},\, z^{(1)},\, \lambda,\, p,\, A^{(1)},\, A^{(2)},\, \mathbf{J}^{(1)},\, \mathbf{J}^{(2)}\right) = \left(1,\, 1,\, 1,\, \mathbf{0},\, 10^{-6},\, 1,\, 1,\, I_3,\, I_3,\, \mathbf{0},\, \mathbf{0}\right).
\end{align*}
The ideal gas equation of state parameters are the same for both the reactants and products of the viscous gas, and given in Table \ref{table:EOSparams}. The heat of combustion is $\mathcal{Q} = 21$. 

Similarly to the viscous detonation problem \ref{sect:viscousdetonation}, there are no analogues for $c_s$ and $c_t$ in the model used by Clarke et al. We use the values $c_t = 1.2$ and $c_s = 1$, which have been chosen such that our results qualitatively relate to the ones in \cite{clarke2}. 

Contrary to the assumptions of this work, the viscosity coefficient and heat conductivity used in \cite{clarke2} are not constant, but depend on the material's temperature, through $\mu = \mu_0 \sqrt{T/T_0}, \kappa = \kappa_0 \sqrt{T/T_0}$, where $\mu_0 = \tilde{\mu_0} / \tilde{\mu} = 1.1832$ and $Pr = \gamma C_v \mu_0 / \kappa_0 = 0.72$.
In our formulation, these variables act through the source systems for distortion and heat conduction. As a result, the assumptions of Jackson's semi-analytic ODE solvers \cite{jackson} will no longer hold in the current case. We work around this by using the odeint library of numerical ODE solvers from Boost 1.71.0.0 for this problem. 

Thermal energy is added at the left boundary, at a high power of $q_{w} = \frac{\gamma \tilde{p} \tilde{c}}{Pr (\gamma - 1)}$. This is achieved by adjusting the energy of the ghost cells, in accordance with the Fourier's law of heat conduction,
\begin{align}
-\kappa^{(2)} \frac{\partial T^{(2)}}{\partial x} = q_{w},
\label{eq:Fouriers}
\end{align}
and also adjusting the thermal impulse vector such that 
\begin{align}
\mathbf{J}^{(2)} = \frac{q_{w}}{c_t^{(2)} T^{(2)}}. 
\end{align}
In practice, we solve the non-linear ODE \eqref{eq:Fouriers} analytically to extrapolate temperature in the region outside the boundary, and then update $\mathbf{J}^{(2)}$ accordingly. Reflective conditions are used for all other state variables for the left boundary, whereas transmissive conditions are used on the right. 

As in \cite{clarke2}, this test is run with both an inert and a reactive gas. For the inert case, $\mathcal{Q} = 0$ and no source term is added to the $\lambda$ equation, i.e.\ $\mathcal{K} = 0$. For the reactive case, Arrhenius kinetics are used,  
\begin{align}
\mathcal{K} = - \lambda C \exp\left(-\frac{T_A}{T}\right),
\end{align}
with $C = 11.56$ and $T_A = 20$. Note that caution should be taken when using a temperature-dependent rate law in combination with a constant density ratio assumption for the explosive mixture (see \cite{stewart}). However, for this particular case, the reactants and products are both modelled using the ideal gas equation of state, with the same parameters. Therefore, both the temperature equilibrium condition and the constant-density ratio assumption give the same outcome, $\rho^{(\alpha)} = \rho^{(\beta)} = \rho^{(2)}$. 

The numerical results for the inert and reactive cases use a non-dimensional spatial resolution of $\Delta x = 0.5$ and are presented in Figure \ref{fig:heatdetonation}. 
By heating the inert gas, a shock wave is formed at the wall, and propagates steadily, while the temperature is continuously rising at the wall. This wave is also produced when heating the reactive gas, as seen in the temperature plots. However, the temperature increase accompanied with the wave is not sufficiently high to detonate the gas. The reaction zone therefore follows the precursor shock, forming a deflagration wave. The reaction wave accelerates, and eventually catches up with the leading shock (at around $t = 104$). From then onwards, the two waves propagate together, progressing towards a steady state in the shape of a ZND detonation. Even if the detonation has not yet converged, it is worth comparing its speed at final times, $2029 \,\si{\meter \per \second}$, with the theoretical value calculated using the Hugoniot curve and Rayleigh line, $D_{CJ} = 1926 \,\si{\meter \per \second}$. 

\subsection{Detonation confined by hydrodynamic material}
\label{sect:hydroconfined}
\begin{figure}[!tb]
	\centering
	\captionof{table}{Initial conditions (non-dimensional) for the hydrodynamically confined detonation problem}
	\label{table:hydro_confined}
	\begingroup
	\setlength{\tabcolsep}{11.7pt} 
	\renewcommand{\arraystretch}{1.1} 
	\scriptsize
	\begin{tabular}{l l c c c c c c c}
		\hline\\[-7pt]
		Material & Region & $\rho^{(1)}, \rho^{(\alpha)}, \rho^{(\beta)}$ & $\mathbf{u}$ & $z^{(1)}$ & $\lambda$ & $p$ & $A^{(1)}, A^{(2)}$ & $\mathbf{J}^{(1)}, \mathbf{J}^{(2)}$ \\[2pt] \hline \\[-6pt]
		LX-17 (Booster) & $[0, 0.1]$ & 1 & $\mathbf{0}$ & $10^{-6}$ & 0 & $0.3$ & $I_3$ & $\mathbf{0}$ \\
		LX-17 & $(0.1, 1.5]$ & 1 & $\mathbf{0}$ & $10^{-6}$ & 1 & $0.001$ & $I_3$ & $\mathbf{0}$\\
		Confiner & $(1.5, 2]$ & 1 & $\mathbf{0}$ & $1 - 10^{-6}$ & 0 & $0.001$ & $I_3$ & $\mathbf{0}$\\
		\hline
	\end{tabular}
	\endgroup
	
	\vspace{10pt}
	
	\includegraphics[width=\linewidth]{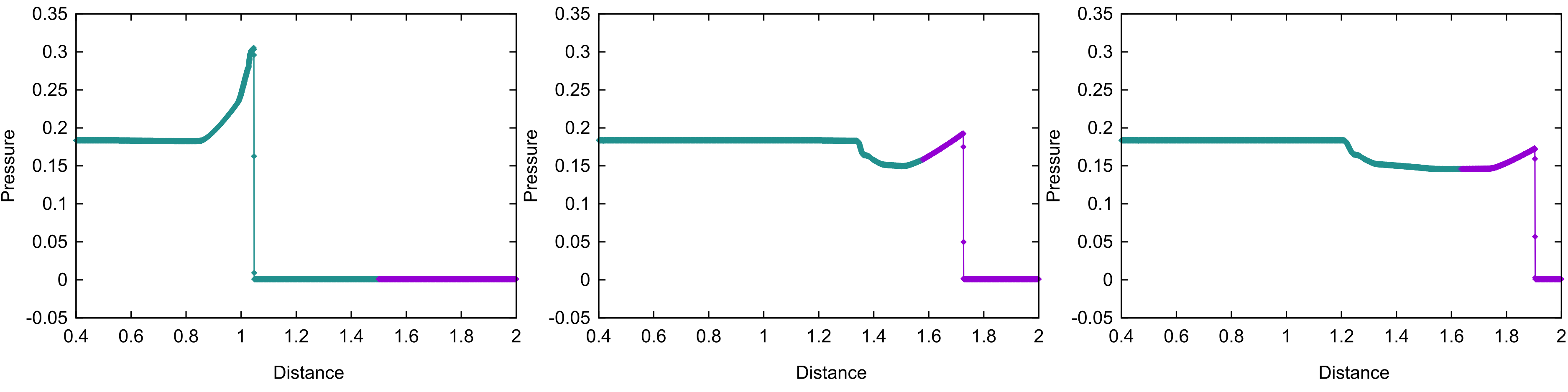}
	\captionof{figure}{Non-dimensional numerical solutions (linespoints) for the test problem with a detonation in LX-17 (green) hitting a hydrodynamic confiner (purple) at times $t = 1.0$ (left), $t = 1.75$ (middle) and $t = 2.0$ (right). (For interpretation of the references to colour in this figure legend, the reader is referred to the web version of this article.)}
	\label{fig:hydro_confined_detonation}
\end{figure}

This test was devised by Michael and Nikiforakis \cite{hybrid} to be the closest one-dimensional setup equivalent to a two-dimensional reactive rate stick problem. The configuration is similar to that of problem \ref{sect:LX-17}, where a detonation forms and propagates through LX-17, but a confiner is also added at the right end of the domain. Therefore, for the solution of this test, our full formulation is reduced to the MiNi16 system. The JWL equation of state \eqref{eq:JWL} is used to model all three materials, and the Ignition \& Growth rate law \eqref{eq:IgnitionGrowth} governs the reaction process. The parameters for both LX-17 and the confiner are summarised in Table \ref{table:EOSparams}. Parameters for the confiner have been taken from \cite{banks08}, and have been re-scaled with respect to the CJ reference state for LX-17 \eqref{eq:LX-refstate}. Both materials are inviscid fluids, so we set $\tau_1 = 0$ and $c_s = 0$. The Ignition and Growth parameters \eqref{eq:IG-LX-17} from the LX-17 shock initiation problem are employed for the reaction process. 

We use a booster of high pressure to initiate the explosive. The initial conditions for the test are given in Table \ref{table:hydro_confined}. The domain is chosen to be large enough to allow the booster-ignited explosive to run to steady state before impacting with the confiner. 

Figure \ref{fig:hydro_confined_detonation} depicts the numerical solution of pressure at times $t = 1, 1.75, 2$. In the first plot, the generated detonation wave travels through the body of the explosive, while the explosive-confiner interface remains stationary. When the detonation wave hits the interface, the peak pressure is seen to significantly decrease. The pressure in the explosive region near the interface is also decreased by the rarefaction wave that decompresses the region. Also, as seen by the position of interface (change in colour of plot), we can deduce that the interaction of the detonation wave and the contact has caused the confiner to comply. 

Even though a different closure condition has been used (constant density ratio in place of temperature equilibrium), our results qualitatively match those of Michael and Nikiforakis \cite{hybrid}. 

\subsection{Detonation confined by elastoplastic material}
\label{sect:elastoplasticconfined}
\begin{figure}[!tb]
	\centering
	\captionof{table}{Initial conditions (non-dimensional) for the elastoplastically confined detonation problem}
	\label{table:elastoplastic_confined}
	\begingroup
	\setlength{\tabcolsep}{10pt} 
	\renewcommand{\arraystretch}{1.1} 
	\scriptsize
	\begin{tabular}{l l c c c c c c c c}
		\hline\\[-7pt]
		Material & Region & $\rho^{(1)}$ & $\rho^{(\alpha)}, \rho^{(\beta)}$ & $\mathbf{u}$  & $z^{(1)}$ & $\lambda$& $p$ & $A^{(1)}, A^{(2)}$ & $\mathbf{J}^{(1)}, \mathbf{J}^{(2)}$ \\[2pt] \hline \\[-6pt]
		LX-17 (Booster) & $[0, 0.1]$ & 4.6877 & 1 & $\mathbf{0}$ & $10^{-6}$ & 0 & $0.3$ & $I_3$ & $\mathbf{0}$\\
		LX-17 & $(0.1, 1.5]$ & 4.6877 & 1 & $\mathbf{0}$  & $10^{-6}$ & 1 & $0.001$ & $I_3$ & $\mathbf{0}$\\
		Confiner & $(1.5, 2]$ & 4.6877 & 1 & $\mathbf{0}$ & $1 - 10^{-6}$ & 0& $0.001$ & $I_3$ & $\mathbf{0}$\\
		\hline
	\end{tabular}
	\endgroup
	\vspace{10pt}
	\includegraphics[width=0.8\linewidth]{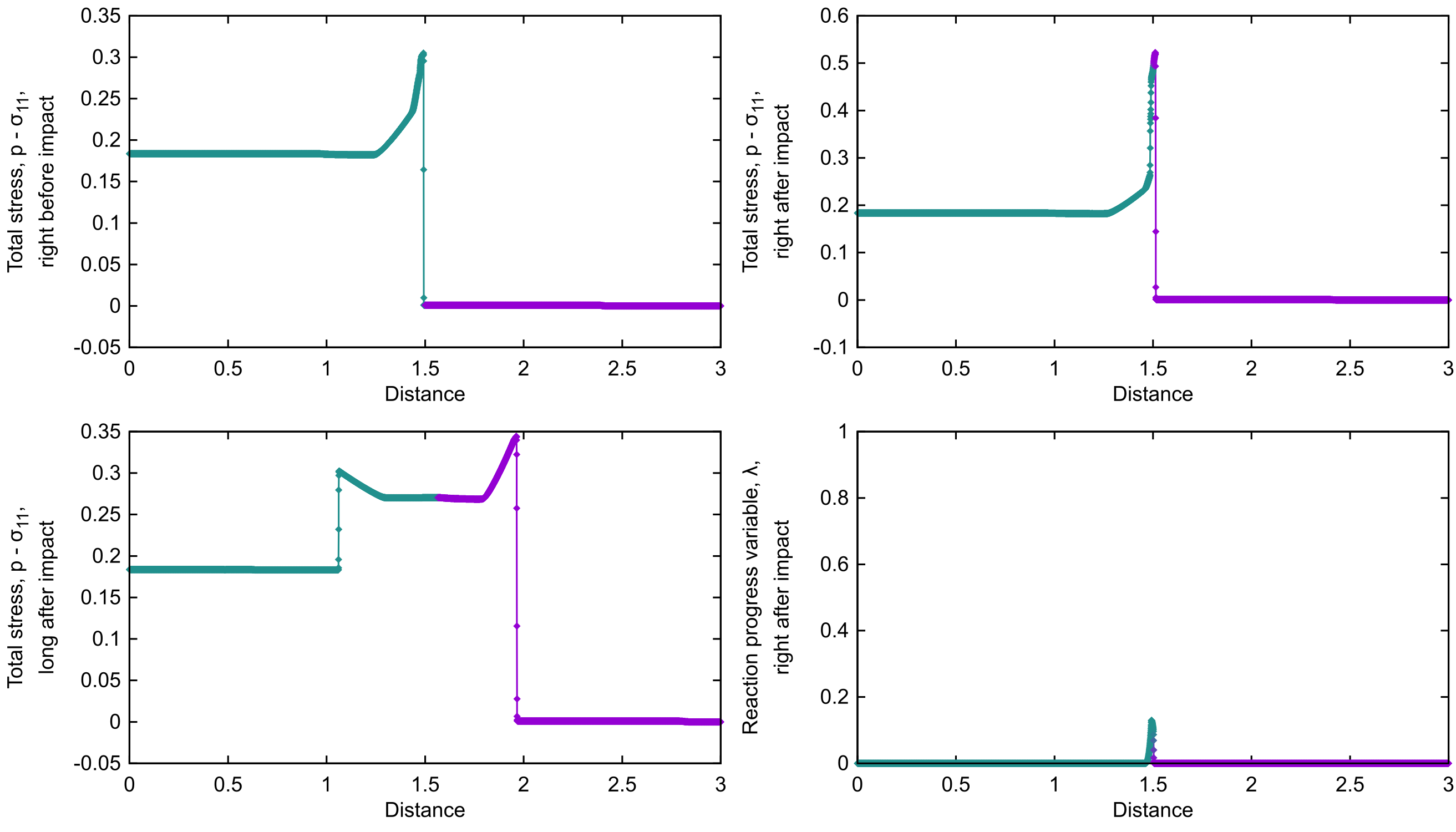}
	\captionof{figure}{Non-dimensional numerical solutions (linespoints) for the test problem with a detonation in LX-17 (green) hitting elastoplastic copper (purple). The stress tensor profiles are plotted at $t = 0, 0.025, 0.675$ after impact, with an additional reaction volume fraction plot at $t = 0.025$ after impact. (For interpretation of the references to colour in this figure legend, the reader is referred to the web version of this article.)}
	\label{fig:elastoplastically_confined_detonation}
\end{figure}

Having considered a number of problems involving fluid, solid and explosive materials, we move on to combine a condensed-phase explosive with an elastoplastic solid. This problem is specifically devised to verify that the developed formulation can cope with the scenario of an elastoplastically confined detonation. We keep the configuration of the previous problem in \ref{sect:hydroconfined}, and replace the previously hydrodynamic confiner with unstressed, elastoplastic copper. The full formulation is therefore required for this problem. LX-17 is used for the explosive, with the same scaled equation of state and reaction rate law parameters as before. Copper is described by the Shock Mie-Gr\"uneisen equation of state,
\begin{equation}
\begin{alignedat}{2}
p_{ref} (\rho) &= \frac{\rho_0 \rho c_0^2 (\rho - \rho_0)}{[\rho - s(\rho - \rho_0)]^2}, \qquad \qquad \,\,\,\, T_{ref}(\rho) &&= 0,\\
e_{ref} (\rho)&= \frac{p_{ref}(\rho)}{2 \rho \rho_0} (\rho - \rho_0), \qquad \qquad \qquad \Gamma(\rho) &&= \Gamma_0 \left(\frac{\rho_0}{\rho}\right)^{N+1}.
\end{alignedat}
\label{eq:hugoniot}
\end{equation}
Its parameters are taken from \cite{peshkov} and scaled with respect to the CJ reference state for LX-17 \eqref{eq:LX-refstate}, to be rendered non-dimensional. They can be found in Table \ref{table:EOSparams}. The shear sound speed for copper is $c_s = 0.29818$. 

Plasticity is incorporated through the distortion tensor relaxation source term. Specifically, a power-law is taken for the characteristic time for strain dissipation, $\tau_1$, as in equation \eqref{eq:tau1}, with parameters $\tau_0 = 10^6, \sigma_0 = 8.01\times 10^{-4}, n = 100$. 

The initial conditions for the test are given in Table \ref{table:elastoplastic_confined}. A non-dimensional numerical resolution of $\Delta x = 0.0005$ is used to produce the solutions in Figure \ref{fig:elastoplastically_confined_detonation}. The last stress profile before the impact can be seen in the top left plot. The two plots on the right show the normal stress tensor component and reaction progress variable, $\lambda$, at a short time after impact. At this time, the explosive is not fully burned, as can be seen from the non-zero $\lambda$ plot. The impact of the detonation wave at the confiner interface causes pressures as high as $0.5222 (58.67\,\si{\giga \pascal})$. Finally, the bottom left plot shows the stress profile at a later time, illustrating clearly the reflected and transmitted waves in copper and LX-17 products. It is important to mention that, as in the hydrodynamically confined detonation problem of section \ref{sect:hydroconfined}, the confiner also complies in this problem. However, as expected from the fact that the confiner is now a solid material, the propagation speed of the material interface is much slower than that of section \ref{sect:hydroconfined}. 

Despite using different material and equations of state, great qualitative agreement is achieved with the similar simulation done by Schoch et al.\ \cite{schoch}. 

\subsection{Heating-induced detonation confined by hydrodynamic material}
\label{sect:heatdetonation_confined}
\begin{figure}[!tb]
	\centering
	\captionof{table}{Initial conditions (non-dimensional) for the hydrodynamically confined heat-induced detonation problem}
	\label{table:heatdetonation_confined}
	\begingroup
	\setlength{\tabcolsep}{11.7pt} 
	\renewcommand{\arraystretch}{1.1} 
	\scriptsize
	\begin{tabular}{l c c c c c c c c}
		\hline\\[-7pt]
		\makecell{Material/\\ Region} & $\rho^{(1)}$ & $\rho^{(\alpha)}, \rho^{(\beta)}$ & $\mathbf{u}$  & $z^{(1)}$ & $\lambda$& $p$ & $A^{(1)}, A^{(2)}$ &  $\mathbf{J}^{(1)}, \mathbf{J}^{(2)}$\\[2pt] \hline \\[-6pt]
		\makecell{Reactive gas/\\$[0, 250]$} & 849.91 & 1 &  $\mathbf{0}$& $10^{-6}$ & 1  & $1$ & $I_3$ & $\mathbf{0}$\\
		\makecell{Water/\\$(250, 500]$} & 849.91 & 1 & $\mathbf{0}$ & $1 - 10^{-6}$ & 0 & 1 & $I_3$ & $\mathbf{0}$\\
		\hline
	\end{tabular}
	\endgroup
	
	\vspace{10pt}
	
	\includegraphics[width=0.8\linewidth]{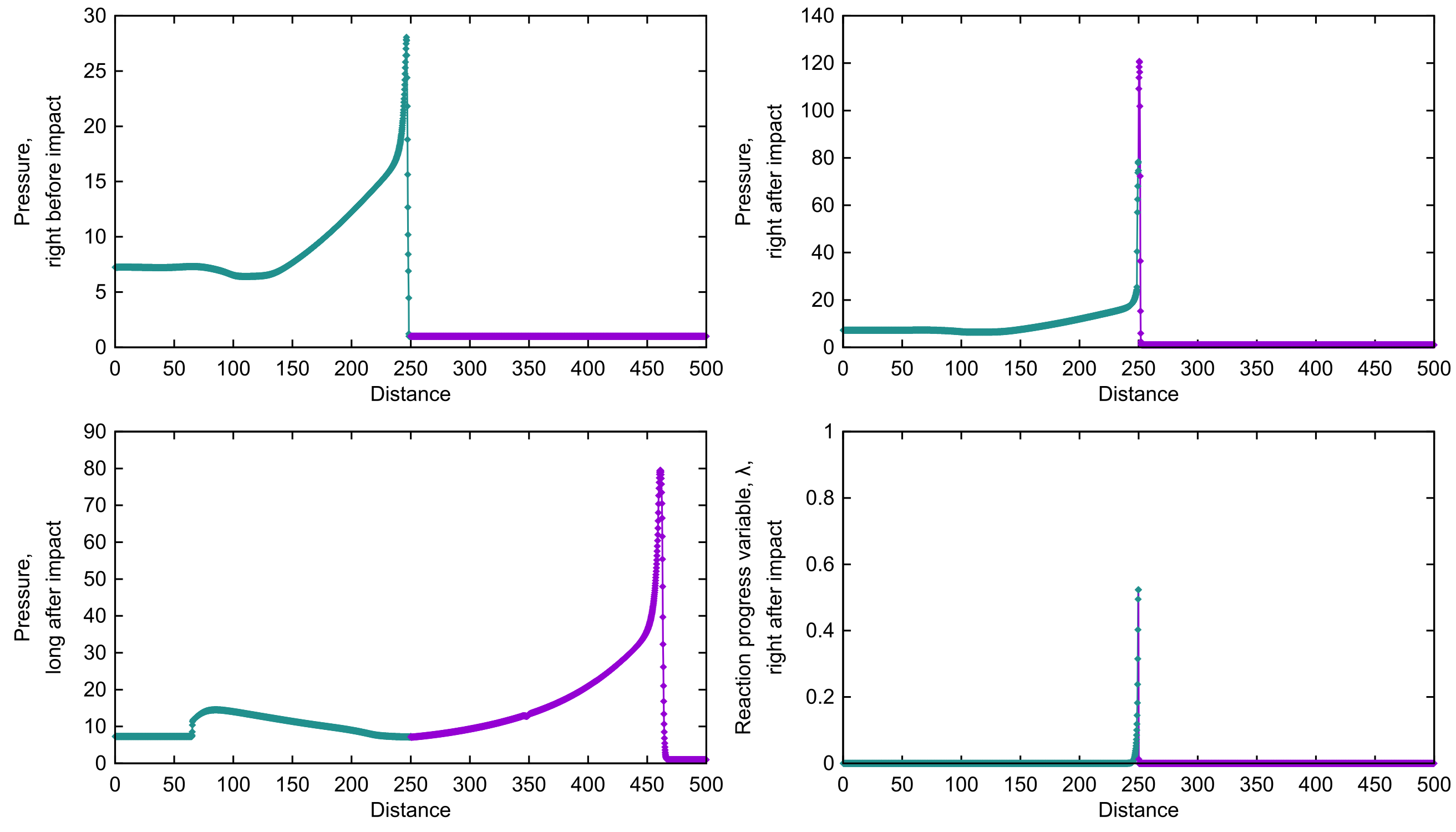}
	\captionof{figure}{Non-dimensional numerical solutions (linespoints) for test problem with a heat-induced detonation in a gas (green) hitting water (purple). The pressure profiles are plotted at $t = 0, 0.5, 38.5$ after impact, with an additional reaction volume fraction plot at $t = 0.5$ after impact. (For interpretation of the references to colour in this figure legend, the reader is referred to the web version of this article.)}
	\label{fig:heatdetonation_confined}
\end{figure}
Finally, we consider the impact of a heating-induced detonation with a hydrodynamic confiner material. This test aims to validate the full multi-material formulation in the reactive, viscous and heat-conducting regime. The initial configuration is similar to problem \ref{sect:heatdetonation}, where a heat source at the left wall initiates a detonation wave in a reactive gas. However, a hydrodynamic confiner (water) is now added to the right end of the domain. The reactive gas is modelled using the ideal gas equation of state, using the parameters of problem \ref{sect:heatdetonation}. The reaction rate law also follows \ref{sect:heatdetonation}, except the activation temperature which we reduce to $T_A = 15$, in order to speed up the ignition process. Similar to problem \ref{sect:heatdetonation}, all variables and parameters used in this problem have been scaled by the reference state \eqref{eq:heatdetonationrefstate} and are hence non-dimensional. Water is described by the stiffened gas equation of state
\begin{equation}
\begin{alignedat}{2}
p_{ref}(\rho) &= - \gamma p_\infty, \qquad \qquad T_{ref}(\rho) &&= 0,\\
e_{ref}(\rho) &= 0,\qquad \qquad \qquad \quad \,\,\Gamma(\rho) &&= \gamma - 1,
\end{alignedat}
\label{eq:SG}
\end{equation}
with parameters given in Table \ref{table:EOSparams}. Being an inviscid fluid, we set $\tau_1 = 0$ and $c_s = 0$ for water. 

The non-dimensional initial conditions can be found in Table \ref{table:heatdetonation_confined}. The initial configuration is such that the initiated detonation wave converges to steady state before reaching the material interface. Furthermore, a higher (non-dimensional) resolution of $\Delta x = \frac{1}{6}$ is used in this problem, to ensure that the reaction zone is sufficiently captured. 

The results at three different times are depicted in Figure \ref{fig:heatdetonation_confined}. The top left plot shows the last pressure profile before impact of the detonation wave with the material interface. This is at steady state, propagating with speed of $6.6$, or $1937 \,\si{\meter \per \second}$ in dimensional terms, which compares very well with the theoretical value of $D_{CJ} = 1926 \,\si{\meter \per \second}$. The two plots on the right show the pressure and reaction progress variable, $\lambda$, at a short time after the impact. At this time-frame, the reactive gas is not fully burned, as can be seen from the non-zero $\lambda$ plot. Right after the impact, peak pressure reaches values as high as 122 (approximately $12 \,\si{\mega \pascal}$). Finally, the reflected and transmitted waves at a long time after the impact can be seen in the bottom left plot of Figure \ref{fig:heatdetonation_confined}. 

As demonstrated in this problem, the full formulation can cope with confined heat-induced detonation problems, which comprise reactive fluid mixtures, immiscible interfaces, as well as viscous and heat conducting behaviour. 

\section{Conclusion} 
\label{sect:concl}
In this work we present a unified formulation for the simultaneous simulation of condensed-phase explosives and structural response. The proposed model builds upon two existing formulations; the MiNi16 model and the GPR model. 

The new formulation is developed to satisfy a number of specifications summarised at the beginning of this paper. It is therefore important to review the full model against these requirements.  Firstly, it should be able to model condensed-phase explosives. The formulation uses the MiNi16 framework to describe reactive flows, thereby inheriting all the relevant properties. Distinct equations of state are allowed for the reactants and products, therefore the reaction zone can be accurately captured. The model's capabilities in the ignition and detonation regimes are assessed in one-dimensional tests for the explosives LX-17 and C-4 (see \ref{sect:LX-17} and \ref{sect:C4}). 

The interaction between explosive and inert materials is also handled by the MiNi16 framework. We demonstrate this by a one-dimensional test involving a detonation in LX-17 hitting an inert fluid confiner material (see \ref{sect:hydroconfined}). 

The modelling of material response is another important requirement for the proposed formulation. This is achieved by using the GPR model, which is a unified model for continuum mechanics. We effectively `insert' a GPR-type material in each of the two phases of the MiNi16 framework. This way, depending on the distortion relaxation time of each phase, we can model viscous and inviscid fluids, as well as elastic, plastic and elastoplastic solids. By choosing appropriate combinations for the relaxation times, the full model reduces to a fluid-fluid, solid-fluid or a solid-solid model. 

The physical process of heat conduction is inherited from the extension of Dumbser et al.\ \cite{dumbser} to the GPR model. The unified nature of the new formulation enables the modelling of mechanically or thermally induced, inviscid or viscous detonations in materials exhibiting structural response and heat conduction. In this work, we consider a viscous shock-induced detonation in section \ref{sect:viscousdetonation}, and a heating-induced detonation in section \ref{sect:heatdetonation}. 

Finally, by using the diffuse interface method for two-phase flows embedded in the MiNi16 model, we are able to simulate interactions between different materials under a single formulation. This in turn implies that the same numerical algorithm can be used over the whole domain. In this work, we use a splitting method to solve the hyperbolic system and source system. We derive an HLLD-type approximate Riemann solver for the spatial terms, which is extended to second order of accuracy using the MUSCL-Hancock method. The relaxation source terms are integrated by adapting the semi-analytic solver of Jackson \cite{jackson} to our multi-phase formulation. We use two one-dimensional tests to demonstrate the capabilities of the full model; a detonation wave in LX-17 hitting quiescent elastoplastic copper in section  \ref{sect:elastoplasticconfined}, and a thermally induced detonation wave in a viscous gas hitting water in section \ref{sect:heatdetonation_confined}. 

Future work will include extending the formulation to incorporate more inert and reactive materials as well as multiple dimensions, to address industrial applications in combustion modelling. 

\setlength{\bibsep}{3pt plus 0.5ex}
\bibliographystyle{elsarticle-num-names}
\bibliography{references}

\end{document}